\documentclass[sigconf]{acmart}
\pdfoutput=1
\usepackage{booktabs} % For formal tables
\usepackage{multirow}
\usepackage{subfigure}
\usepackage{algorithm}
\usepackage{algorithmic}
\usepackage{diagbox}
\begin{document}
\title{An Evaluation of BBR and its variants}
\titlenote{Produces the permission block, and
  copyright information}

\author{Songyang Zhang}
\affiliation{
 \institution{School of Computer Science and Engineering, Northeastern University, China}
}
\email{sonyang.chang@foxmail.com}
%\author{Weimin Lei}
%\affiliation{
%  \institution{School of Computer Science and Engineering, Northeastern University, China}
%}
%\email{leiweimin@ise.neu.edu.cn}
%\author{Wei Zhang}
%\affiliation{
%  \institution{School of Computer Science and Engineering, Northeastern University, China}
%}
%\email{zhangwei1@mail.neu.edu.cn}
%\author{Yunchong Guan}
%\affiliation{
%  \institution{School of Computer Science and Engineering, Northeastern University, China}
%}
%\email{y.c.guan@foxmail.com}
\begin{abstract}
The congestion control algorithm bring such importance that it avoids the network link into severe congestion and guarantees network normal operation. Since The loss based algorithms introduce high transmission delay, to design new algorithm simultaneously achieving high throughout and low buffer occupation is a new working direction. The bottleneck bandwidth and round trip time (BBR) belongs such kind, and it has drawn much attention since its release. There are other algorithms modified from BBR to gain better performance. And the implementation of BBR v2.0 is released recently. We implement a framework to compare the performance of these algorithms in simulated environment.
\end{abstract}

\maketitle

\section{Introction}
Ever since the congestion collapse in internet was observed in 1986, Jocobson \cite{Jacobson1988Congestion} proposed to implement the additive increase and multiplicative decrease to regulate the congestion window of TCP flow and the research works on congestion control seem endless. The congestion control for TCP bears such importance that it guarantees the internet can work in normal status and avoids congestion collapse happening again. Several algorithms have been proposed by making minor changes to the basic AIMD control law and to adapt it in some specific network environments \cite{Lin2019Extensive}\cite{Turkovic2019Fifty}.

As the development of the internet infrastructure, it is realized that these rate control algorithms based on AIMD do not perform well. These AIMD-like algorithms taking packet loss as link congestion indication tend to fill the pipe. Especially in today’s internet, large buffers are configured in intermediate routers and the notorious bufferbloat problem \cite{Gettys2011Bufferbloat}  is introduced. It is reported in \cite{Cardwell2016BBR}, users experience delays of seconds to minutes in cellular network.

In recently, developing congestion control algorithms to achieve high throughout while minimizing transmission delay at the same time becomes a new trend. BBR \cite{Cardwell2016BBR} is one typical example of such kinds. Since its release by google in 2016, BBR has gain much attention. Google first deployed BBR in its B4 wide area network and YouTube. Several works \cite{Hock2017Experimental, Scholz2018Towards, Jain2018Design} were published to analyze its behavior either in simulated links or in real network testbed.

It was concluded that BBR can significantly improve throughput for TCP connection from these experiments with different sources. For such merit, BBR is widely applied to build VPN (virtual private network) service for rate acceleration purpose. BBR suffers from RTT unfairness issue (reported in \cite{Hock2017Experimental}\cite{Ma2017Fairness}) and tends to overload the bottleneck link when flows coexisting, which causes considerable packets loss in links with shallow buffer. 

There are some works Tsunami\footnote{https://github.com/dlxg/Linux-NetSpeed\label{tsunami}}, BBRPlus\footnote{https://github.com/cx9208/bbrplus\label{bbrplus}}, BBR+\cite{Wang2019Active} to make some modification to these control parameters in BBR in order to gain better performance or apply it to a different network domain. Both Tsunami and BBRPlus have gain some attention. BBRPlus has won 623 stars and 280 forks on github. Whether the two are applied in real program are not clear. There are no public available report to analyze their performance too. In May 2019, google has released BBR v2 in QUIC \footnote{https://www.chromium.org/quic} codebase with a goal to better coexistence with Reno and Cubic \cite{Ha2008CUBIC} flows. And its implementation in Linux net stack can be got at \footnote{https://github.com/google/bbr/blob/v2alpha/net/ipv4/tcp\_bbr2.c}.

In this article, a framework is implemented and the performance of BBR algorithm and its variants (Tsunami, BBRPlus, BBR+ and BBR v2) are evaluated on simulated network environment. Since the release of BBR v2 is recent event, it seems no other works doing a full evaluation on its performance. 

The transmission protocol implemented in simulation is a simplified version of QUIC protocol and only three frames (STREAM, STOP WAITTING and ACK) are applied to build a reliable transmission protocol on top UDP. The framework is about 10000 lines c++ codes  and it can be available at \footnote{https://github.com/SoonyangZhang/DrainQueueCongestion}. The collected data trace during simulation is about 1Gbps after compression and  is public available at \footnote{https://pan.baidu.com/s/170pReidrkW7TzTBar6wKBA: q9m2}

The rest of this paper is organized as follows. The background and related works on congestion control are briefed in Section 2. The detail on these algorithms are presented on Section 3. Section 4 is the simulation results and evaluation. The conclusion and discussion is made on Section 5.
\section{Related work}
In October of 1986, the data rate from LBL to UC Berkeley dropped from 32kbps to 40bps, which was recorded as the first congestion collapse event in internet \cite{Jacobson1988Congestion}. A series mechanisms e.g. slow start, round trip time variance estimation and the AIMD control rule were introduced to achieve network stability. These ideas are implemented in TCP Reno and it remained as the default congestion control algorithms in FreeBSD and Linux . The transport connection should obey the packet conservation principle when network system in stability: a new packet can be injected into the network only when an old packet leaves.

In Reno, the packet loss event is interpreted as network congestion signal. On every RTT, Reno sender could inject one more packet into
network to probe more available bandwidth and multiplicatively reduces congestion window size by half when packet loss happens to recover the network back to normal status. Its adjustment on congestion window ($w$) can be summarized as Equation \eqref{eq:aimd}. For Reno, $\alpha=1$ and $\beta=0.5$.
\begin{equation}
\label{eq:aimd}
w(t+1)=
\begin{cases}
w(t)+\frac{\alpha}{w(t)},& \text{when ack is received}\\
\beta w(t),& \text{when packet loss is detected}
\end{cases}
\end{equation}

Based on the fluid model \cite{Misra2000Fluid}, a differential equation on the rate adjustment process of a AIMD flow can be deduced as Equation \ref{eq:fluid}. $p$ is packet loss rate. Let $\dot x=0$, the throughout at equilibrium is $x=\frac{1}{RTT}\sqrt{\frac{1-p}{p}}$, which is usually used as rate control on these flows to be friendly to AIMD flow.
\begin{equation}
\label{eq:fluid}
\dot x=\frac{1-p}{{RTT}^2}-\frac{x^2p}{2}
\end{equation}

The reason to choose AIMD as congestion window control is it can guarantee bandwidth allocation fairness \cite{Chiu1989Analysis} with a decentralized solution. Kelly \cite{Kelly1998Rate} modeled the congestion control algorithm as optimization problem by introducing utilization function. The goal is to maximum the service satisfactory to each user under the constraint of the capacity ($c_l$) of bottleneck link. $x_s$ denotes the sending rate of user $s$, $U(x_s)$ measures the satisfaction or welfare of user $s$.
\begin{equation}
\label{eq:utility}
\begin{aligned}
\max \quad & \sum_{s}U(x_s)\\
\textrm{s.t.} \quad &\sum_{s\in S(l)}x_s\leq c_l\\
\end{aligned}
\end{equation}

To find the optimal solution in a global view is impractical in consideration of larger scale and heterogeneity of real networks. A showdown price in introduced to decompose the primal problem into its dual form as shown. A decentralized solution is shown in Equation \eqref{eq:shadow}. $p_l$ can be interpreted as price per unit bandwidth at link $l$. Such rate iteration is a mathematical expression to AIMD control law. The rate adjustment in Equation \eqref{eq:shadow} is to converge to the optimal value by iteration. For the first time, the flow rate control problem is backup by mathematical theory. The utility maximization theory is applied for stability analysis in new designed rate control algorithms and is further developed in \cite{Low1999Optimization}.
\begin{equation}
\label{eq:shadow}
\dot x_s(t)=\lambda(w_s-x_s(t)\sum_{l\in L(s)}{p_l})
\end{equation}

In high speed network, once a packet loss happens, the classic AIMD algorithm will take quite long time to recover back to the congestion window before the multiplicative reduction action and has a low utilization of bandwidth resource. Several algorithms e.g. STCP\cite{Kelly2003Scalable}, HSTCP\cite{Leith2004H}, BIC\cite{LisongXu2004Binary}, Cubic\cite{Ha2008CUBIC} have been proposed to remedy such problem. The congestion window increase and decrease behavior is modified to adapt for high speed network in these solutions. STCP and HSTCP introduce several RTT unfairness, with which the flow with short RTT obtain more bandwidth. In BIC, the congestion control is viewed as a binary search problem to enable aggressive bandwidth probe. The congestion window grows to the middle point of $w_{min}$ and $w_{max}$. $w_{max}$ is the congestion window before the last fast recovery and $w_{max}$ is the window value after the fast recovery. When the packet loss is detected, the middle point is assigned to $w_{max}$, and it is taken as the new value of $w_{min}$ otherwise. Cubic is an update version of BIC. A cubic function is introduced for window adjustment. It can be solve the RTT unfairness issue since its congestion window increase is only depended on the time between two consecutive congestion events. The Cubic algorithm achieves better performance than AIMD, and it has been the default configuration in Linux net stack until now.

There are algorithms taking delay as congestion signal. The delay based algorithms can prevent queue from building up, make highly use of channel resource and maintain throughput stability. Once the delay has exceeded of some threshold, the congestion window will be reduced to alleviate congestion. Vegas \cite{Brakmo1995TCP} and Fast \cite{Wei2006FAST} are belonging to such kind. Even Vegas can achieve quite low queue occupation, it get starvation when sharing links with loss based algorithms. For such reason, it is not widely applied in real network. There are other protocols TCP-LP \cite{Kuzmanovic2003TCP} and LEDBAT \cite{Rossi2010LEDBAT} following the working mechanism of AIMD and taking delay as congestion signal. These two take themselves junior to TCP flows and actively yield bandwidth to when sharing links with high priority TCP flows.

Some other algorithms take both delay and packet loss to indicate link congestion. Veno \cite{ChengPengFu2003TCP} takes backlog queue delay to differential random loss and congestion loss to improve TCP throughput performance in wireless network. In Illinois \cite{Liu2008TCP}, the window adjustment parameters $\alpha$ and $\beta$ are dynamically changed with delay. When delay signal is small, a large value of $\alpha$ is applied for fast convergence. Compound \cite{Tan2006Compound} is the rate control algorithm in Windows operating systems. The congestion widow is increased fast when bandwidth resource is available.

Attempts are made to the coexistence of delay based congestion control with delay base congestion control. When facing packet loss, CHD \cite{David2010Improved} will back off the congestion window with probability. When the backlog queue delay is above $q_{th}$, the possibility to reduce congestion window decreases in CHD to gain better competiveness with buffering filling flows. CDG \cite{Hayes2011Revisiting} used delay gradient to infer link buffer in full, empty, rising, and falling status. CDG has the ability to tolerate non-congestion related loss and only backs off for congestion related packet loss.

Given the delay problem introduced by loss based algorithms, to design congestion control with high throughput and low delay becomes trend. Serval works e.g. Sprout \cite{Winstein2013Stochastic}, PCC \cite{Dong2015PCC}, BBR \cite{Cardwell2016BBR} and Copa \cite{Arun2018Copa} were designed with such porpose.

Sprout, Verus \cite{Zaki2015Adaptive}, ExLL \cite{Park2018ExLL} and C2TCP \cite{Abbasloo2019C2TCP} are mainly designed for cellular network. Legacy algorithms have degenerate performance in links with highly variable capacities and non-congested packet loss. DCTCP \cite{Alizadeh2010Data} and Timely \cite{Mittal2015TIMELY} are proposed to achieve low message delivery delay and high throughout for datacenter networks. Some works e.g. Remy \cite{Winstein2013TCP}, QTCP \cite{Li2018QTCP} and TCP-Drinc \cite{Xiao2019TCP} apply reinforce learning in rate control. They are evaluated in simulated paltform, whether such methods can be applied in network stack need further investigation. 

As the popular of real time communication applications, deploying rate control algorithms on multimedia traffic is a necessary to avoid congestion and to promote fair bandwidth allocation. GCC \cite{Carlucci2017Congestion}, NADA \cite{Zhu2018NADA}, and SCReAM \cite{Johansson2017Self} are optimized for interactive multimedia transmission. GCC is the default rate control algorithm in WebRTC \footnote{https://webrtc.org/} project, which enables video communication among browsers. 
\section{Algorithm detail}
\begin{figure}
\includegraphics[height=2.5in, width=3in]{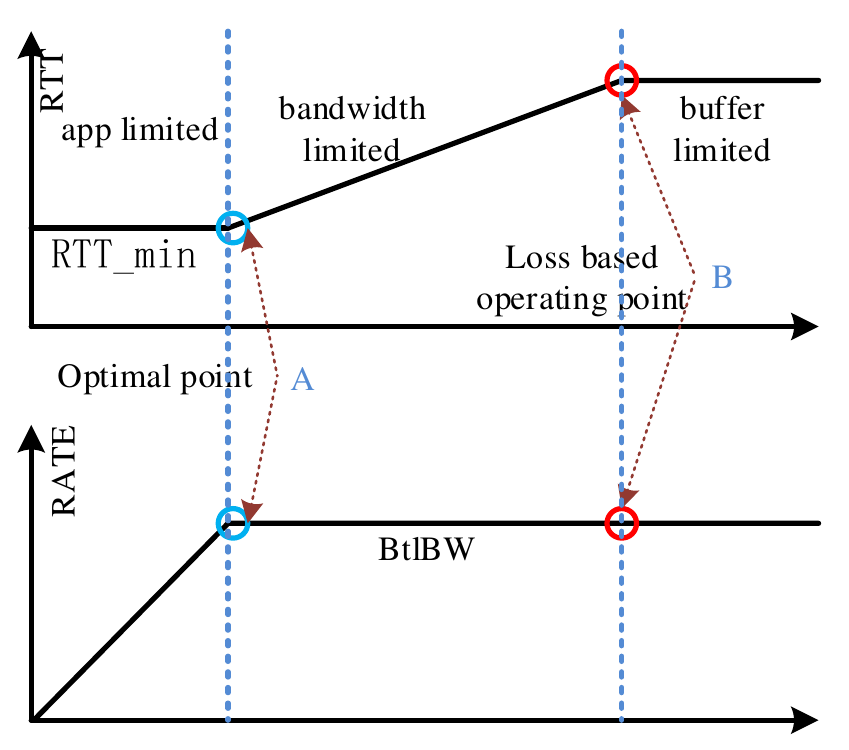}
\caption{Congestion control operating point, based on \cite{Cardwell2016BBR}}
\label{Fig:operating}
\end{figure}
Loss based algorithms were designed to avoid congestion, but it inevitably leads the network into overuse. The capacity of a bottleneck link is denoted as $BtlBW$. When only one flow is presence in this link and it packet sending rate is less than $BtlBW$, the packet round trip delay is the minimum $RTT_{min}$ as shown in Figure \ref{Fig:operating}. But the flow with loss based rate control will keep increase its rate when no packet loss happens. The rate finally exceeds $BtlBW$ and these extra sent packets will be buffered at routers. The packets can be received with the full rate $BtlBW$ at the cost of increased delay. The points on left edge of the bandwidth limited region achieve the same throughout but lower delay compared with loss based operating point. The optimal point in congestion control is named after Kleinrock \cite{Kleinrock1979Power}, maximizing throughput while minimizing delay and loss. However, the Kleinrock’s point can not be reached with distributed algorithm, proved in \cite{Jaffe1981Flow}. BBR is proposed to get close to the optimal point.
\subsection{BBR}
In BBR, the bandwidth can be estimated on each received acknowledge packet. When a new packet is sent out, the packet state information ($total\_byte\_acked, last\_acked\_packet\_ack\_time$) is recorded. $total\_byte\_acked$ counts for the bytes that successfully received by its peer. $last\_acked\_packet\_ack\_time$ is the received time of last acknowledgement packet. When a $sent\_packet$ is acknowledged at $now$, a new bandwidth estimation sample can be calculated as Equation \eqref{eq:bwes}. The $BW$ of the channel is the maximum $bw\_es$ within 10 RTTs. And the minimal rtt $RTT_{min}$ is monitored during the whole phase.
\begin{equation}
\begin{aligned}
&last\_acked\_packet\_ack\_time=now\\
&total\_byte\_acked+=sent\_packet.bytes\\
&\Delta t=now-sent\_packet.last\_acked\_packet\_ack\_time\\
&\Delta delivered=total\_byte\_acked\\
&-sent\_packet.total\_byte\_acked\\
\end{aligned}
\label{eq:deliver}
\end{equation}
\begin{equation}
\label{eq:bwes}
bw\_es=\frac{\Delta delivered}{\Delta t}
\end{equation}

\begin{figure}
\includegraphics[height=1in, width=3in]{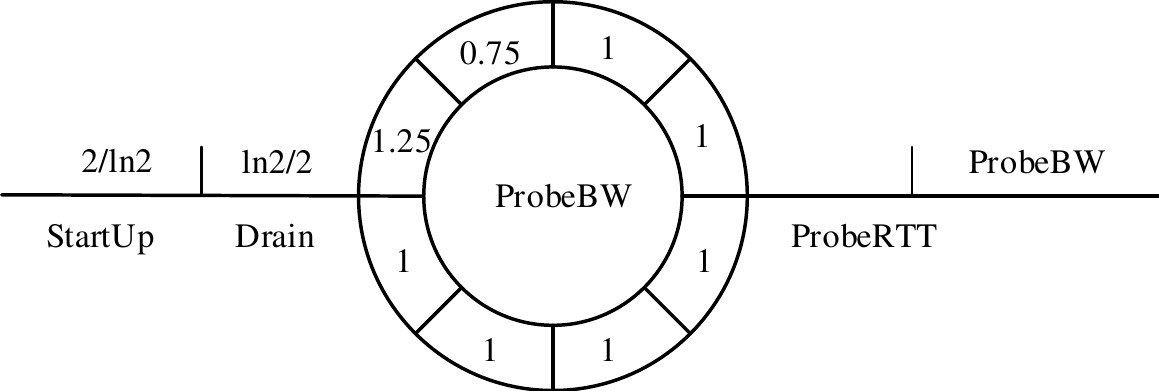}
\caption{Control states in BBR}
\label{Fig:state}
\end{figure}

There are four control states StartUp, Drain, ProbeBW, and ProbeRTT in BBR as shown in Figure \ref{Fig:state}. In each control state, the packet sending rate ($pacing\_rate$) is the product of $pacing\_gain$ and $BW$.

The states StartUp and Drain are applied at session initial phase in BBR. The startup state is quite similar to slow start in TCP. In StartUp, $pacing\_gain$ is $\frac{2}{ln2}$ to double the inflight packets at each RTT to let sender probes the maximum available bandwidth. When newly estimated bandwidth is 1.25 times less than the previous value and such circumstance lasts
for 3 times, the pipe seems to be get fully filed and the state is changed into Drain state. The $pacing\_gain$ in Drain is $\frac{ln2}{2}$ to decrease the sending rate below $bw$. Until the inflight packets match $BDP$ ($BW*RTT\_{min}$), state is changed from Drain to ProbeBW. 

During ProbeBW state, The $pacing\_gain$ cycles in 8 RTTs with different values ($kPacingGain[]=[1.25, 0.75, 1, 1, 1, 1, 1, 1]$). The probe up phase with 1.25 gain is to increase the sending rate to probe more available bandwidth, and probe down phase with 0.75 gain is to get rid of the excess queue which may be accumulated in the probe up phase.

The $cwnd$ is set as $2*BDP$ in ProbeBW to guarantee enough packets can be sent during probe up phase. If $cwnd$ is set exactly equal with $BDP$, the acknowledgement clocking is affected by disturbance in real network, new packets can not be sent out during probe up once the $cwnd$ is exhausted and a small estimated bandwidth is got. If $RTT_{min}$ is not sampled again within 10 seconds, the link is deemed falling into congestion, and ProbeRTT state id applied. The $cwnd$ is set as 4*MSS. Until the inflight packets size is less than 4*MSS, new packets are injected into the network and ProbeRTT will last at most 200 milliseconds. In ProbeRTT, the inflight packets are nearly totally drained from links and a new $RTT_{min}$ value is sampled.

There are two different versions to update the $pacing\_gain$ from probe down phase to probe cruise phase in ProbeBW. The first is to increase the cycle offset when the prove down phase holds for essentially 1 $RTT_{min}$. The second is to increase cycle offset only when the inflight packets Less than or equal to the target $BDP$ to achieve lower queue delay than the first form. Most published works only analyse the performance of BBR in the first form. In later part, both versions will be evaluated. For convenience, the second form is named as BBR'.
\subsection{BBRPlus}
BBRPlus was first introduced in a blog \footnote{https://blog.csdn.net/dog250/article/details/80629551}. In the origin version of BBR, the duration of probe down lasting 1 $RTT_{min}$ introduces considerable latency. When the network system changes a lot, the fixed probe cycle length may not adapt well. The probe cycle length is randomized from 2 to 8 (line 4 in Algorithm \ref{alg:plusgain}) in BBRPlus. The probe down phase exists only when the inflight packets match the estimated $BDP$ (line 11-12 in Algorithm \ref{alg:plusgain}). Such change is to improve fairness and reduce packet loss when multiple flows sharing a bottleneck. 

The procedure to update $pacing\_gain$ in BBRPlus is shown in Algorithm \ref{alg:plusgain}. $kGainCycleLen$ is 8 and $CYCLE\_RAND$ equals 7. When there is packet loss event, the probe up phase exists earlier (line 14 in Algorithm \ref{alg:plusgain}).
\begin{algorithm}[htb] 
\caption{UpdateGainCyclePhase} 
\label{alg:plusgain} 
\begin{algorithmic}[1]
\REQUIRE ~~\\ 
the timestamp (now), $inflight$, has\_loss
\STATE $elapsed\gets now-cycle\_mstamp\_$
\IF{$elapsed>cycle\_len\_*RTT_{min}$}
\STATE $cycle\_mstamp\_\gets now$
\STATE $cycle\_len\_\gets kGainCycleLen-rand()\%CYCLE\_RAND$
\STATE $pacing\_gain\gets 1.25$
\RETURN
\ENDIF
\IF{$pacing\_gain==1.0$}
\RETURN
\ENDIF
\IF{$pacing\_gain<1.0 \AND inflight\leq BDP$}
\STATE $pacing\_gain\gets 1.0$
\ENDIF
\IF{$elapsed>RTT_{min} \AND(inflight>1.25*BDP \OR has\_loss)$}
\STATE $pacing\_gain\gets 0.75$
\ENDIF
\end{algorithmic}
\end{algorithm}
\subsection{BBR+ and Tsunami}
The performance of Cubic and BBR is tested over LTE on high speed rail (HSR) in \cite{Wang2019Active}. The authors concluded that BBR achieves suboptimal performance in networking environment where both bandwidth and RTT change rapidly. The bandwidth probe strategy and RTT estimation do not adapt well to the network dynamics in HSR situation. The sequence to update $pacing\_gain$ is set to be more radically as [1.5, 0.5, 1.5, 0.5, 1.5, 0.5, 1.5, 0.5] in BBR+. The constant $RTT_{min}$ is too conservative over the last 10 seconds in BBR, and a compensation is added according to Equation \eqref{eq:compensation} in HSR environment. $\lambda^2$ is the shape parameter of gamma distribution. They observed the traced RTT values approximately follow a shifted gamma distribution with a fat tail. In our simulation, the compensation part on the $RTT_{min}$ is not implemented.
\begin{equation}
\label{eq:compensation}
RTprop=RTT_{min}+\lambda\sqrt{Var(RTT)}
\end{equation}

From the perspective of egoism, Tsunami applies the sequence [1.5, 0.75, 1.25, 1.25, 1.25, 1.25, 1.25, 1.25] for gain value update during ProbeBW state in order to gain higher throughput. It gains 264 stars and 136 forks on github. As its name claims, when there is extra bandwidth resource available, Tsunami may quick occupy it. If when the capacity of the bottleneck is fully occupied, Tsunami will cause quite high packet loss rate. Such rate adjustment behavior is harmful to other flows without benefiting itself. 
\subsection{BBR v2}
\begin{figure}
\includegraphics[height=2.5in, width=3in]{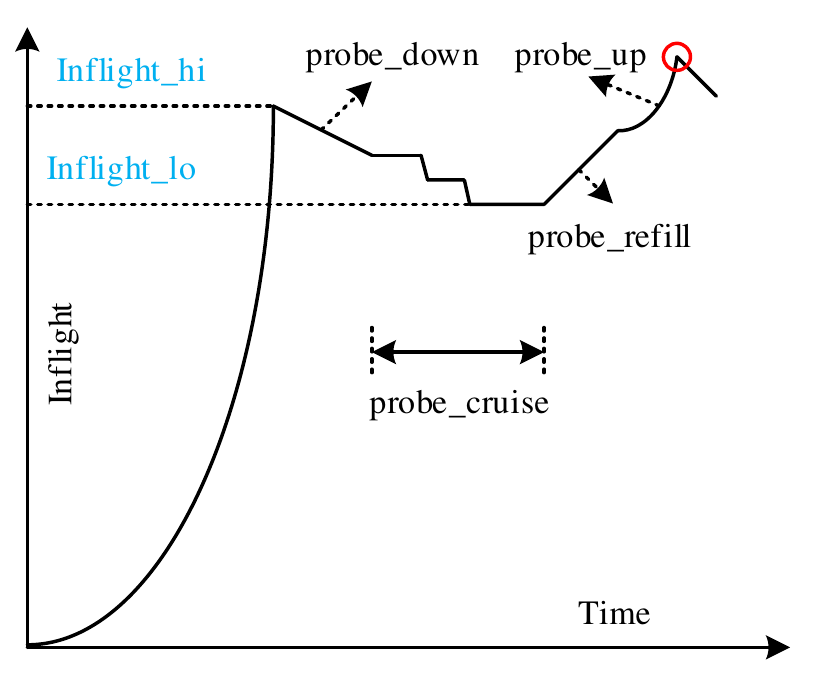}
\caption{BBR v2 flow life cycle, from \cite{Cardwell2019BBR}}
\label{Fig:life-cycle}
\end{figure}
BBR flow will send packets at the full estimated bandwidth and packet loss is not exploited to indicate link congestion. If the queue length at the bottleneck is smaller than $1.5*BDP$ \cite{Cardwell2019BBR}, multiple BBR flows will cause high packet loss. Reno/CUBIC flows gain low throughput when share bottleneck with BBR flows. BBR v2 \cite{Cardwell2019BBR} is proposed to solve these issues in BBR v1. BBR v2 is claimed that it can make better coexistence with Reno/CUBIC flows and achieves low queue delay.

BBR v2 takes packet loss into its control logic. The life cycle of BBR v2 flow is shown in Figure \ref{Fig:life-cycle}. When the estimated bandwidth dost not exceed the target for 3 times, the sender assumes reaching the full bandwidth in BBR v1. Besides that, condition on packet loss is added to exist from StartUp to Drain: the number of loss packet in a round exceeds 8 and packet loss rate exceeds $loss\_threshold$ (0.02). Such condition is applied to avoid excessive packet loss. If the condition on packet loss holds true, the calculated $BDP$ is assigned to $inflight\_hi$. When a new acknowledge packet arrives, $inflight\_lo$ is updated as Equation \eqref{eq:inflightlo}. Here, $\Delta delivered$ is calculated in the same way in Equation \eqref{eq:deliver} and $kBeta$ is 0.3.
\begin{equation}
\label{eq:inflightlo}
\begin{aligned}
&inflight=\Delta delivered\\
&inflight\_lo=\max(inflight,inflight\_lo*(1-kBeta))
\end{aligned}
\end{equation}

In ProbeBW state, the working mechanism of BBR v2 is quite different from BBR v1. The phases (probe\_up, probe\_down, probe\_cruise) switching is no longer depended on the time interval $RTT_{min}$. $cwnd$ is also not set as $2*BDP$ and is related with $inflight\_lo$ and $inflight\_hi$. $inflight\_hi$ is updated if the inflight packets are too high (inflight\_too\_high), in which the loss packet rate exceeds $loss\_threshold$ in last round.
In the probe\_down phase, the $pacing\_gain$ is 0.75, the phase will be switched to probe\_cruise if the inflight packets are drained to $BDP$ or the condition $inflight\_too\_high$ holds true. The $cwnd$ in probe\_cruise is calculated by Equation \eqref{eq:cwndhead}. $kHeadRoom$ is 0.15. $inflight\_hi$ indicates the channel is in dangerous area. To leave headroom to $cwnd$ is to alleviate link congestion to some extent. The $interval$ for probe\_cruise is randomized from 2 seconds to 3 seconds. If duration in probe\_cruise phase exceeds the $interval$, a probe\_refill phase is applied as shown in Figure \ref{Fig:life-cycle}. $cwnd$ is set as $inflight\_hi$ to increase the inflight packets in a round. The probe\_refill is to make preparation for probe\_up.
\begin{equation}
\label{eq:cwndhead}
\begin{aligned}
&inflight\_headroom=inflight\_hi*(1-kHeadRoom)\\
&cwnd=\min(inflight\_lo,inflight\_headroom)
\end{aligned}
\end{equation}

In probe\_up phase, $cwnd$ is increases exponentially per round:1, 2 ,4, 8…. It makes a fast probe to if extra bandwidth available. Once lost bytes are too much, probe\_down phase is applied to get rid of excess queue, as the red ring shows in Figure \ref{Fig:life-cycle}. Algorithm \ref{alg:cwnd-expon} is to exponentially increase $inflight\_hi$ per round. In ProbeRTT, $cwnd$ is reduced by half in v2 to remedy the throughput variation. 
\begin{algorithm}[htb] 
\caption{ProbeInflightHighUpward} 
\label{alg:cwnd-expon} 
\begin{algorithmic}[1]
\REQUIRE ~~\\ 
bytes\_acked, is\_round\_end
\STATE $probe\_up\_acked+=bytes\_acked$
\IF{$probe\_up\_acked\geq probe\_up\_bytes$}
\STATE $delta=\lfloor\frac{probe\_up\_acked}{probe\_up\_bytes}\rfloor$
\STATE $probe\_up\_acked-=delta*probe\_up\_bytes$
\STATE $inflight\_hi\gets inflight\_hi+delta*MSS$
\ENDIF
\IF{$is\_round\_end$}
\STATE $growth\gets 1<<probe\_up\_rounds$
\STATE $probe\_up\_rounds\gets \min(30,probe\_up\_rounds+1)$
\STATE $probe\_up\_bytes\gets \lfloor\frac{cwnd}{growth}\rfloor$
\STATE $probe\_up\_bytes\gets \max(MSS,probe\_up\_bytes)$
\ENDIF
\end{algorithmic}
\end{algorithm}
\section{Evaluation}
\begin{figure}
\includegraphics[height=1.2in, width=3in]{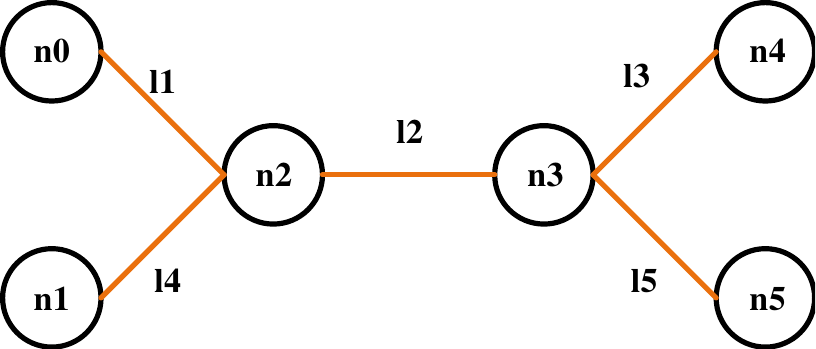}
\caption{Network topology}
\label{Fig:topology}
\end{figure}

These algorithms are evaluated on ns3.26 \footnote{https://www.nsnam.org/} platform. A dumbbell topology as shown in Figure \ref{Fig:topology} is built.
\subsection{Intra protocol fairness}
To test whether this algorithms can guarantee bandwidth allocation property, four flows are created from source $n2$ to destination $n3$. These parameters in Table \ref{tab:l2} to configure link $l2$ are bandwidth (in unit of Mbps), one way propagation delay (in unit of milliseconds) and queue length in nodes. There are total 11 experiments. In each running case, these flows follow a same rate control algorithm. Each simulation process lasts about 400 seconds. The time points to send packets into the network of the four flows are different. The first flow starts at 0 and ends at 400s, the life length of the second flow is 40s to 400s, the third flow is 80s to 200s and the fourth flow is 120s to 300s. At the sender side, when a new packet can be sent out, the rate of the congestion controller is traced. The packet sent time is tagged into ns packet object for receiver to computer one way transmission delay. The one way transmission delay is an indicator to the occupied buffer status in routers. Besides one way delay, at received side, the length of received packet is also recorded.
\begin{table}[]
\centering
\caption{the configuration of $l2$}
\label{tab:l2}
\begin{tabular}{|c|c|c|c|}
\hline
Case & Bandwidth& Propagation delay& Queue length \\ \hline
1 & 5Mbps & 50ms & 5Mbps*100ms \\ \hline
2 & 5Mbps & 50ms & 5Mbps*150ms \\ \hline
3 & 5Mbps & 50ms & 5Mbps*200ms \\ \hline
4 & 6Mbps & 50ms & 6Mbps*100ms \\ \hline
5 & 6Mbps & 50ms & 6Mbps*150ms \\ \hline
6 & 7Mbps & 50ms & 7Mbps*150ms \\ \hline
7 & 7Mbps & 100ms & 7Mbps*300ms \\ \hline
8 & 8Mbps & 100ms & 8Mbps*200ms \\ \hline
9 & 8Mbps & 100ms&  8Mbps*300ms \\ \hline
10 & 10Mbps & 50ms&  10Mbps*150ms \\ \hline
11 & 10Mbps & 50ms&  10Mbps*200ms \\ \hline
\end{tabular}
\end{table} 
\begin{figure}
\centering
\includegraphics[width=3in]{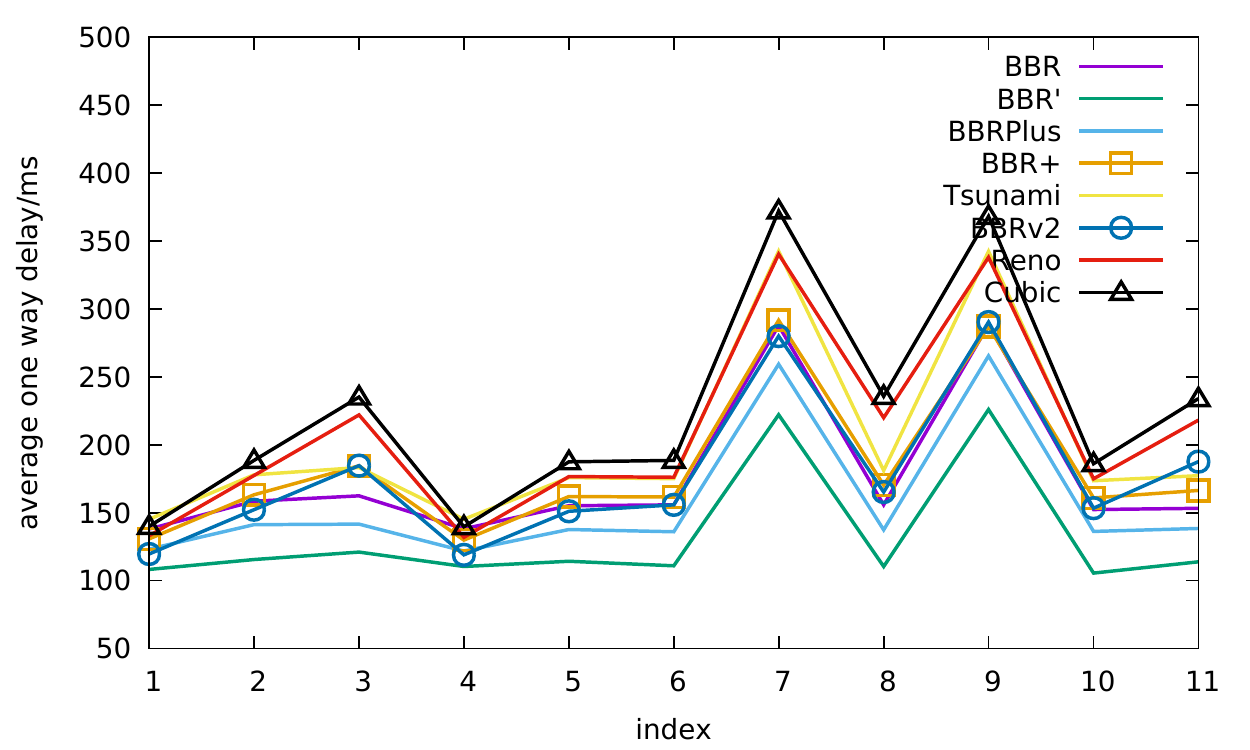}
\caption{Average one way transmission delay}
\label{Fig:owd-comapre}
\end{figure}
\begin{figure}
\centering
\includegraphics[width=3in]{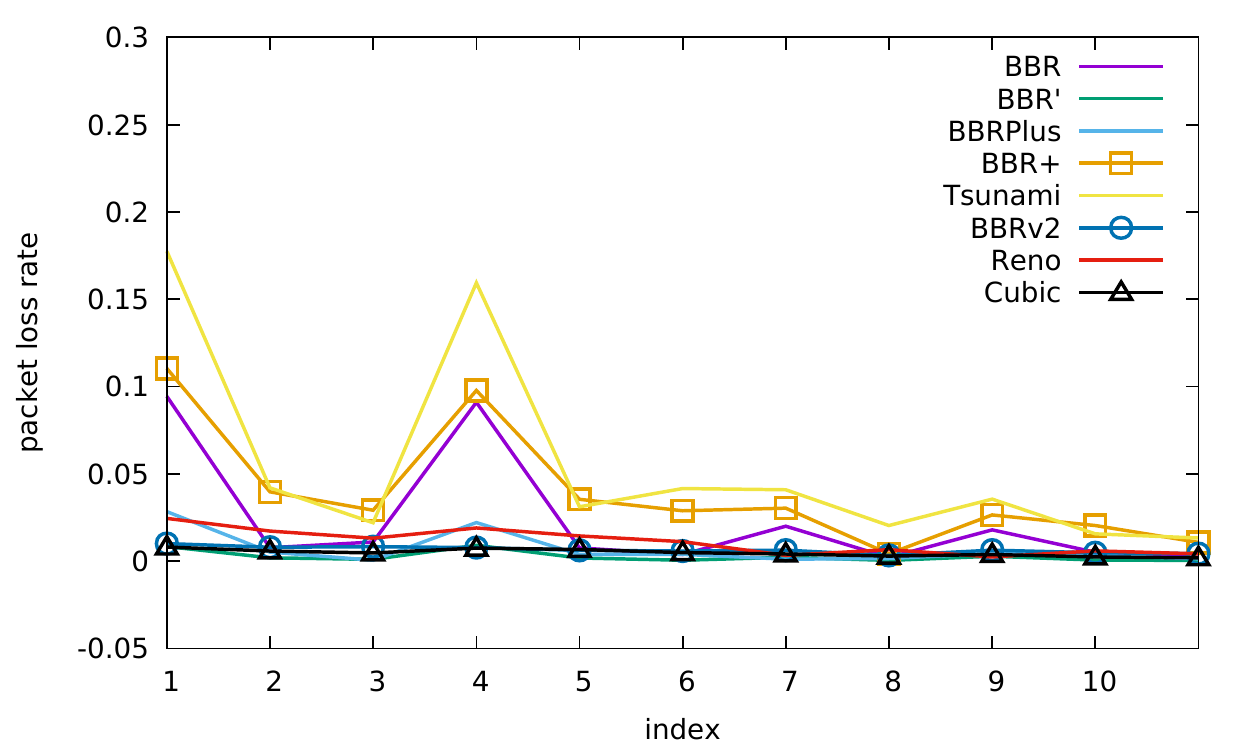}
\caption{Average packet loss rate}
\label{Fig:loss-comapre}
\end{figure}

The average transmission delay of all flows in each experiment is calculated and the results are shown in Figure \ref{Fig:owd-comapre}. The results on average packet loss rate is shown in Figure \ref{Fig:loss-comapre}. Two other buffer filling algorithms Reno and Cubic are also tested.

\begin{figure}[!htb]
\begin{tabular}{cc}
\subfigure[C1]{
\begin{minipage}[t]{0.5\linewidth}
    \includegraphics[width = 1\linewidth]{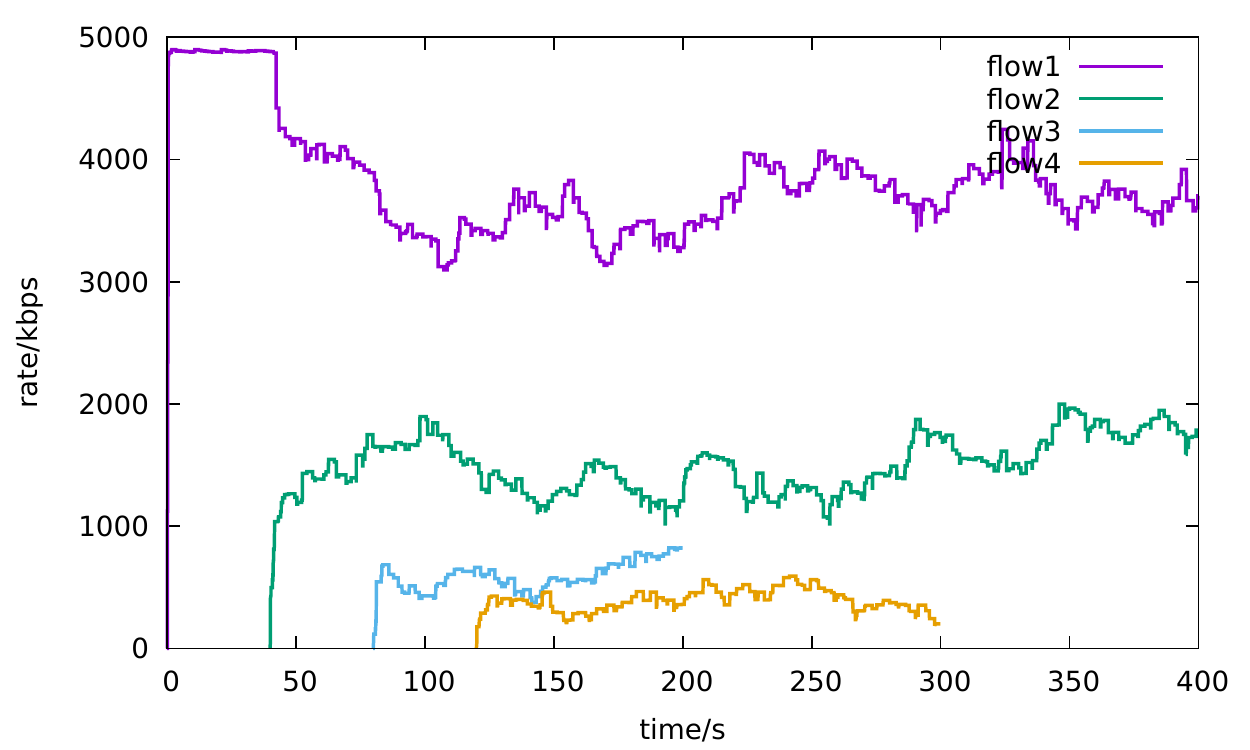}
\end{minipage}}
\subfigure[C3]{
\begin{minipage}[t]{0.5\linewidth}
    \includegraphics[width = 1\linewidth]{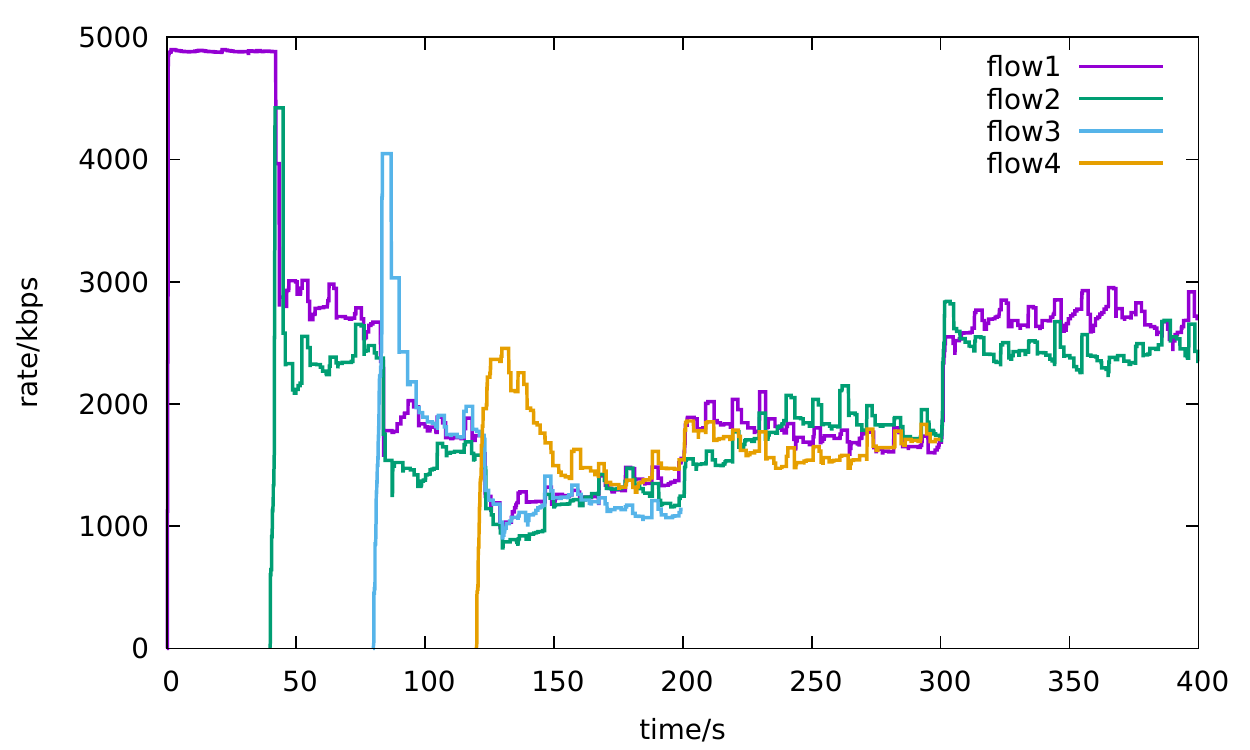}  
\end{minipage}}
\end{tabular}
\caption{Rate dynamics of BBR flows}
\label{Fig:bbr-1-3} 
\end{figure}

Due to the limitation of space, only two cases work as example for further analysis. The link buffer is configured as $1*BDP$ in Case 1 and $2*BDP$ in Case 3. In shallow buffer case, the four BBR flows are not reach the bandwidth fairness line as shown in Figure \ref{Fig:bbr-1-3}(a). With BBR flows presence, there is considerable packet loss rate (about 9\%) in Case 1 as shown in Figure \ref{Fig:loss-comapre}. Such high packet loss rate will impact the bandwidth estimation at sender side. In case 3, the bandwidth allocation fairness is achieved. But when a new flow is initialized, it tends to make over estimation on the bandwidth during the StartUp state, as the steep spike shown in Figure \ref{Fig:bbr-1-3}(b) in flow2 and flow3. Such rate spikes will lead the network into congestion and introduce packet loss. It is the reason for higher packet loss rate (1\%) for Case 3 with longer link buffer compared with Case 2, in which the average packet loss rate is 0.7\%. 

\begin{figure}[!htb]
\begin{tabular}{cc}
\subfigure[C1]{
\begin{minipage}[t]{0.5\linewidth}
    \includegraphics[width = 1\linewidth]{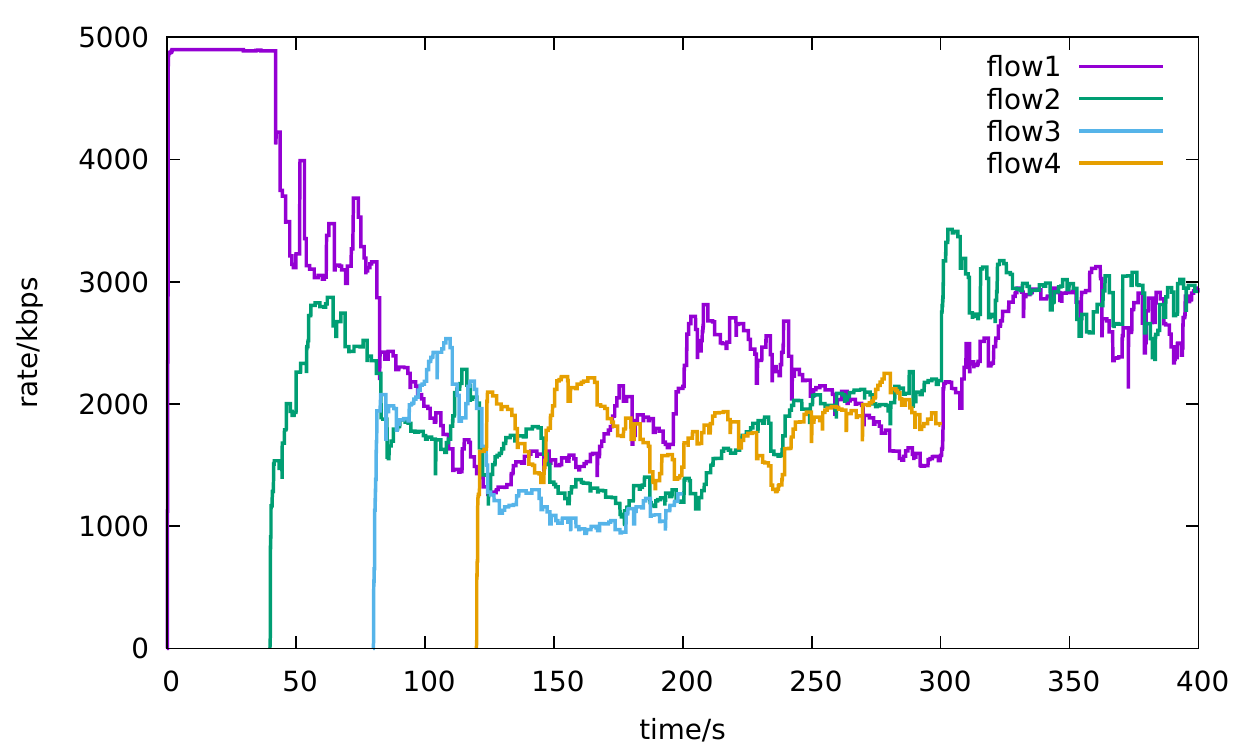}
\end{minipage}}
\subfigure[C3]{
\begin{minipage}[t]{0.5\linewidth}
    \includegraphics[width = 1\linewidth]{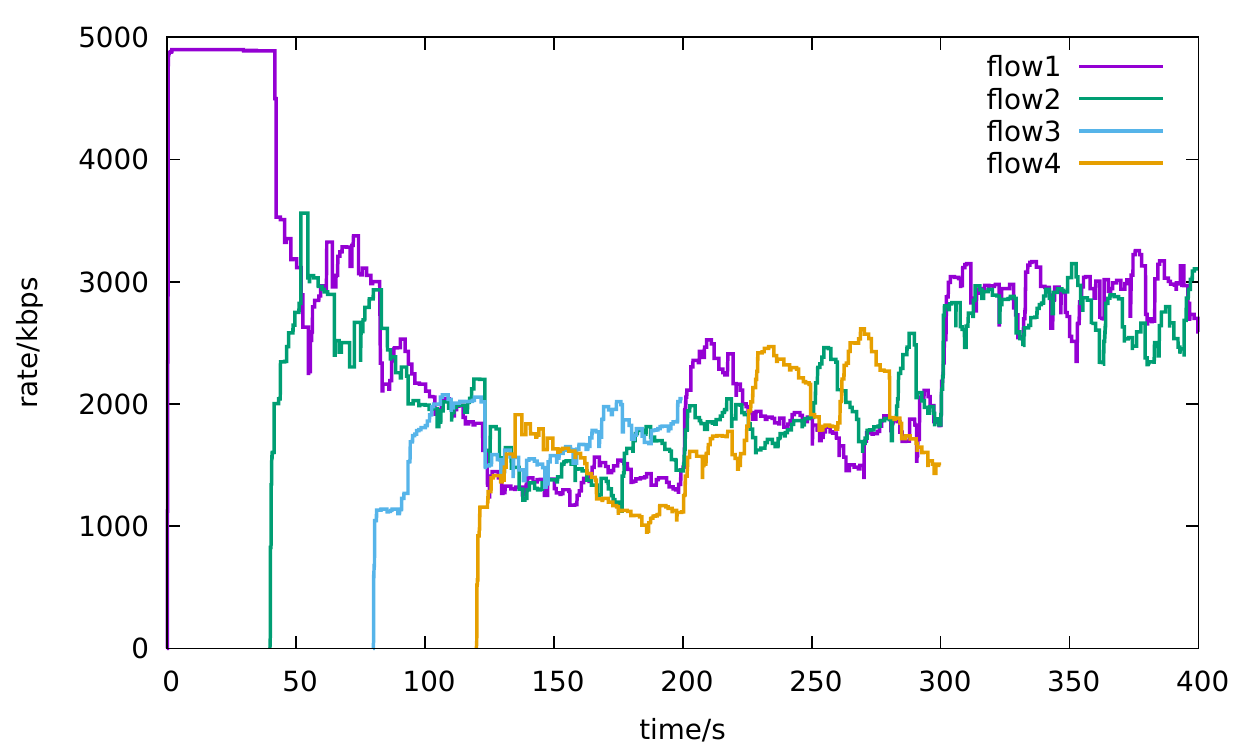}  
\end{minipage}}
\end{tabular}
\caption{Rate dynamics of BBR' flows}
\label{Fig:bbrd-1-3} 
\end{figure}

BBR’ will drain the inflight packets to match the estimated $BDP$ in probe down phase. It can achieve the lowest queue delay as shown in Figure \ref{Fig:owd-comapre} in all tested algorithms, and lower packet loss rate compared with BBR. It also achieve better bandwidth allocation fairness than BBR in shallow buffer case as shown in Figure \ref{Fig:bbrd-1-3}(a). But it introduces rate variation. The rate of BBR’ flow is not quite stable as BBR flow, as shown in Figure \ref{Fig:bbrd-1-3}(b) and \ref{Fig:bbr-1-3}(b).

\begin{figure}[!htb]
\begin{tabular}{cc}
\subfigure[C1]{
\begin{minipage}[t]{0.5\linewidth}
    \includegraphics[width = 1\linewidth]{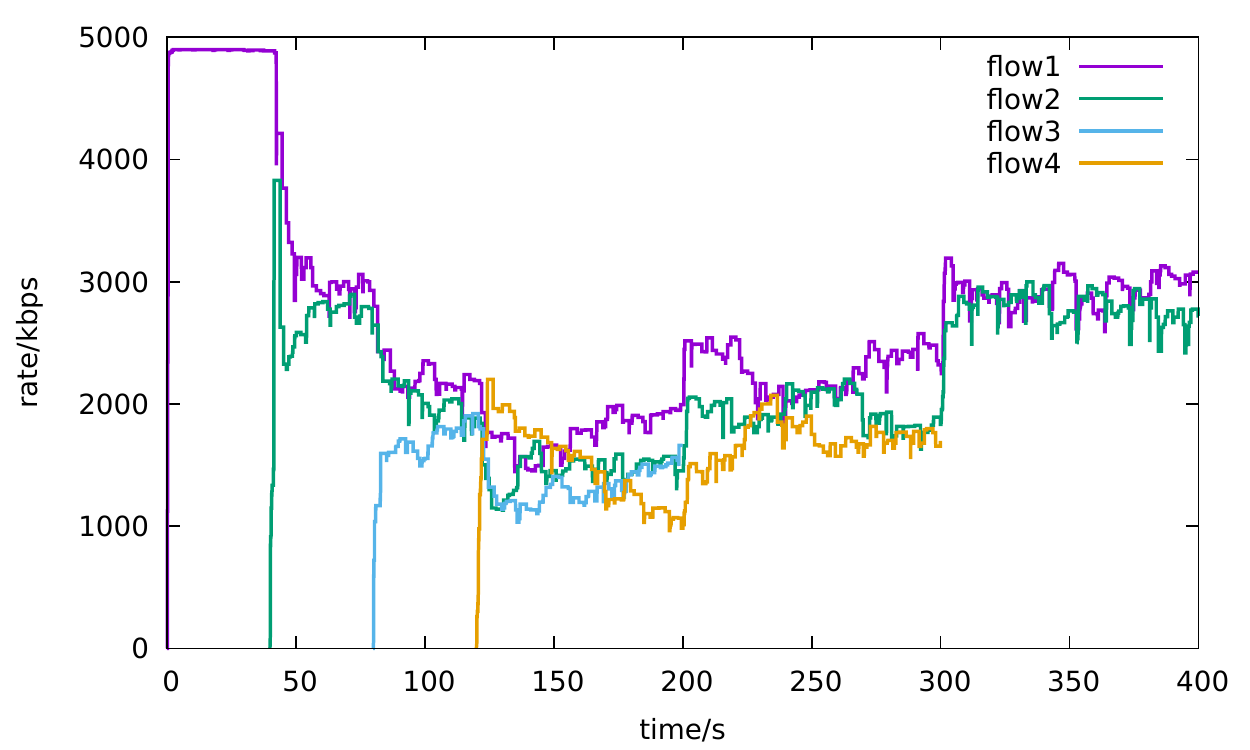}
\end{minipage}}
\subfigure[C3]{
\begin{minipage}[t]{0.5\linewidth}
    \includegraphics[width = 1\linewidth]{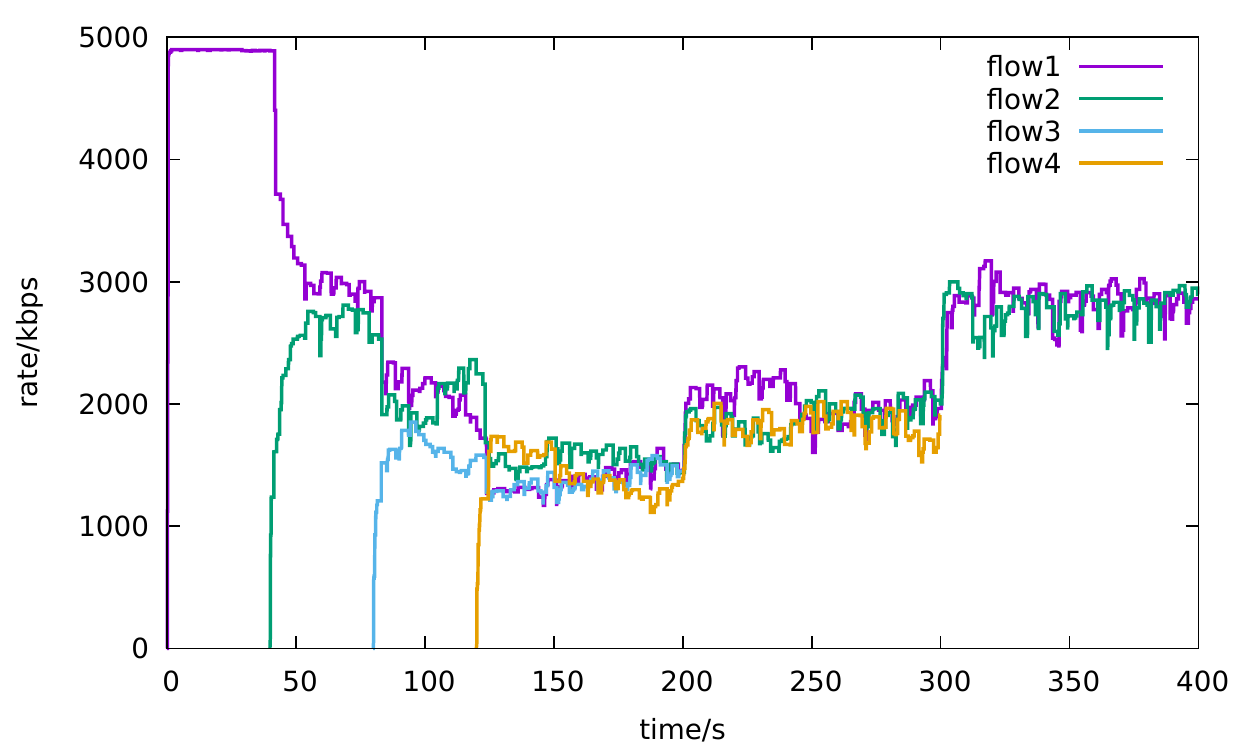}  
\end{minipage}}
\end{tabular}
\caption{Rate dynamics of BBRPlus flows}
\label{Fig:bbrplus-1-3} 
\end{figure}

BBRPlus can achieve the second low average transmission delay. The average packet loss rate is also quite low. At each stage, the throughputs of each flow are quite close. There are some small spikes during the rate adjustment process as shown in Figure \ref{Fig:bbrplus-1-3}.

\begin{figure}[!htb]
\begin{tabular}{cc}
\subfigure[C1]{
\begin{minipage}[t]{0.5\linewidth}
    \includegraphics[width = 1\linewidth]{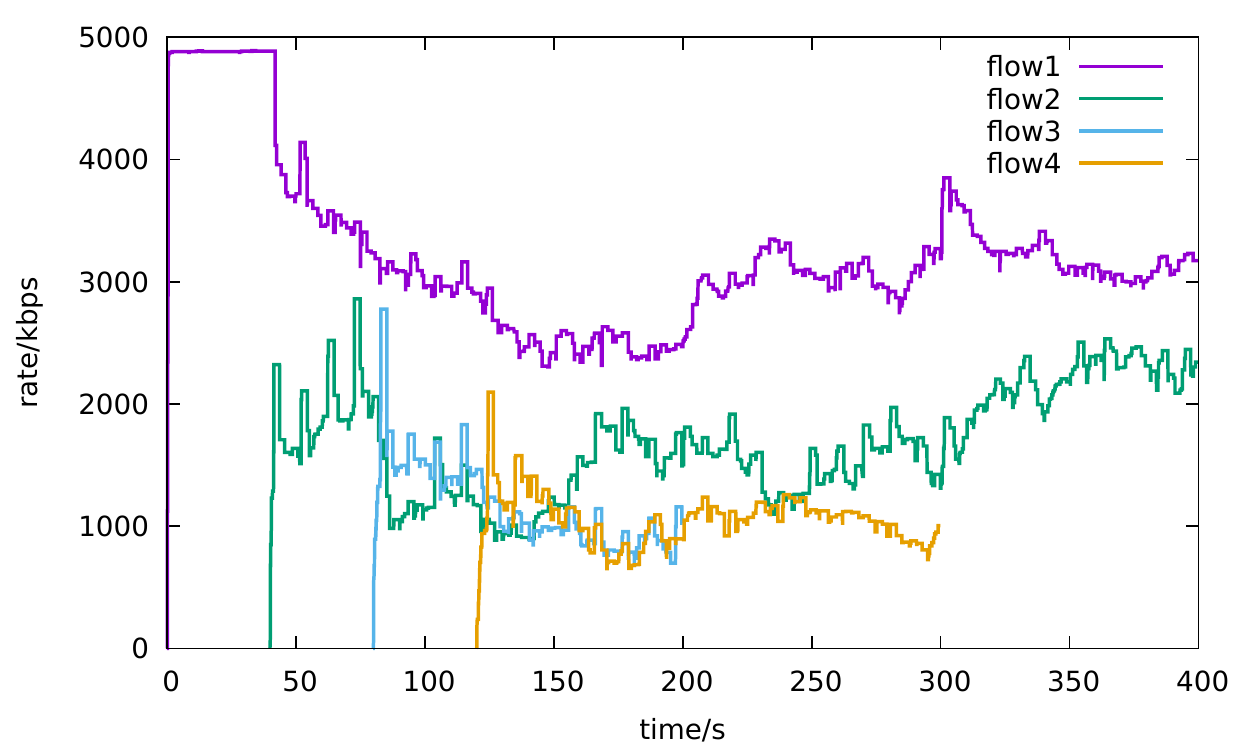}
\end{minipage}}
\subfigure[C3]{
\begin{minipage}[t]{0.5\linewidth}
    \includegraphics[width = 1\linewidth]{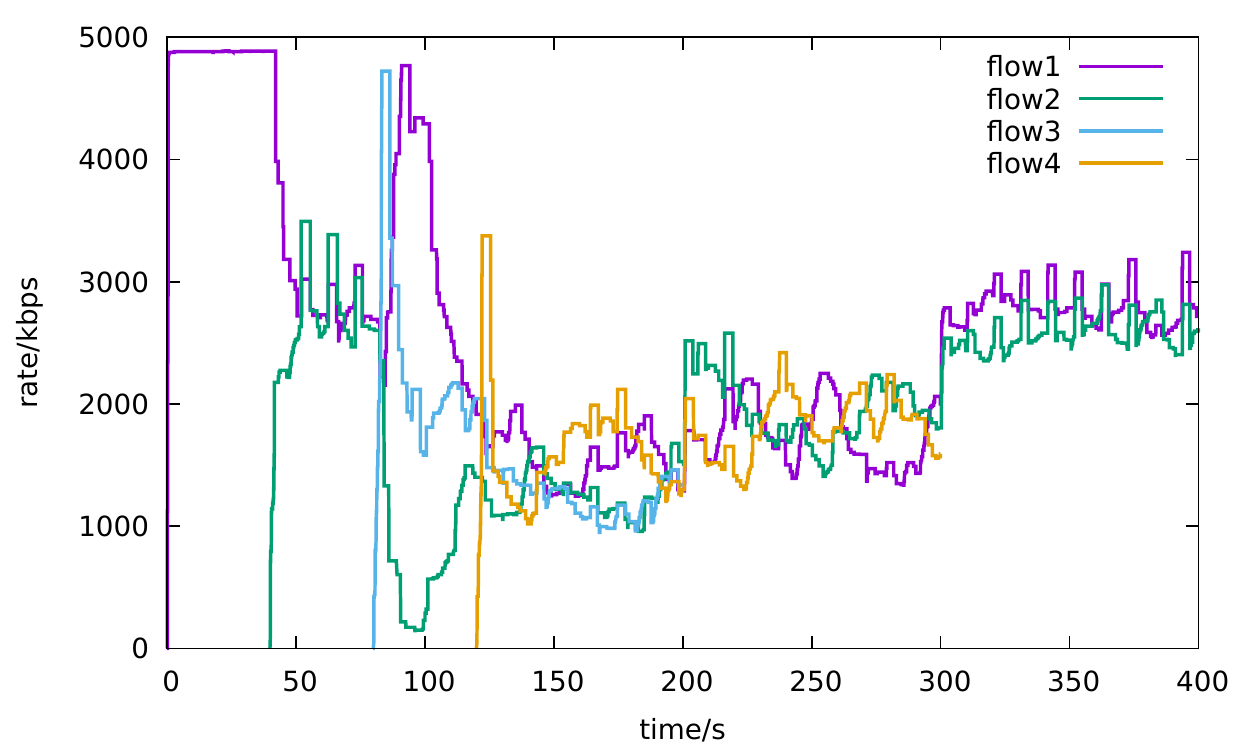}  
\end{minipage}}
\end{tabular}
\caption{Rate dynamics of BBR+ flows}
\label{Fig:hsr-1-3} 
\end{figure}

Both Tsunami and BBR+ suffer from the bandwidth allocation fairness in links with shallow buffer. The way of Tsunami to adjust rate will get the buffer fully occupied. Tsunami flows have the highest packet loss rate and the average transmission delay is quite high. The rate High rate variation can be observed of BBR+ flows in Figure \ref{Fig:hsr-1-3}(b).
\begin{figure}[!htb]
\begin{tabular}{cc}
\subfigure[C1]{
\begin{minipage}[t]{0.5\linewidth}
    \includegraphics[width = 1\linewidth]{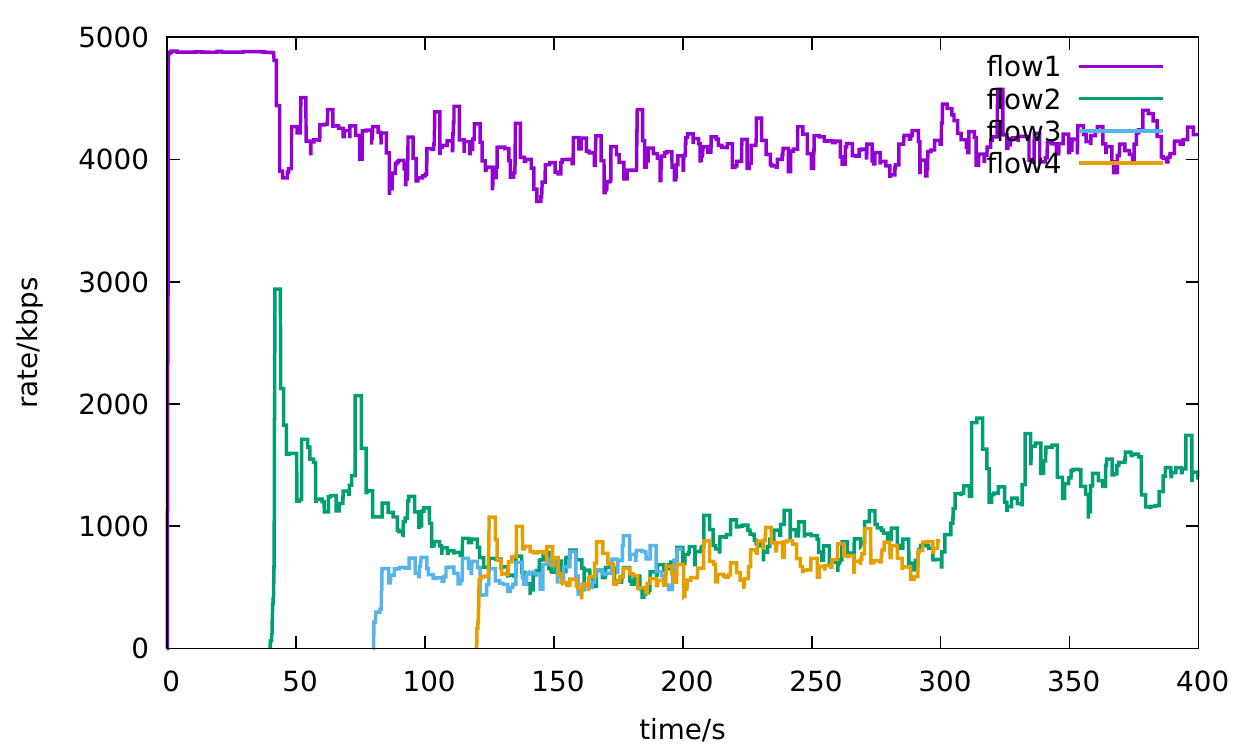}
\end{minipage}}
\subfigure[C3]{
\begin{minipage}[t]{0.5\linewidth}
    \includegraphics[width = 1\linewidth]{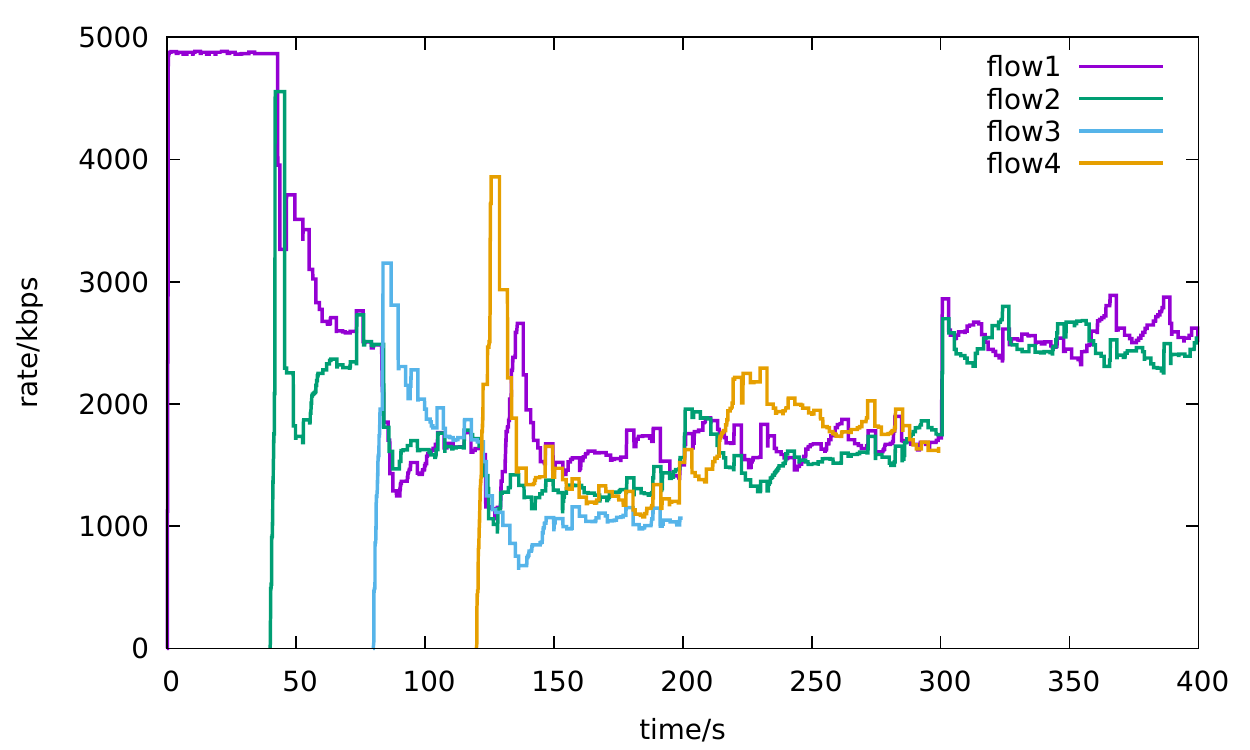}  
\end{minipage}}
\end{tabular}
\caption{Rate dynamics of Tsunami flows}
\label{Fig:tsu-1-3} 
\end{figure}

BBR v2 flows can maintain well bandwidth allocation fairness. The rate adjustment is quite frequency in BBR v2. That is the result of the balance between probing more bandwidth and avoiding link congestion. The lower queue delay in some test cases is lower than BBR but is still higher than BBR’ in all test cases.
\begin{figure}[!htb]
\begin{tabular}{cc}
\subfigure[C1]{
\begin{minipage}[t]{0.5\linewidth}
    \includegraphics[width = 1\linewidth]{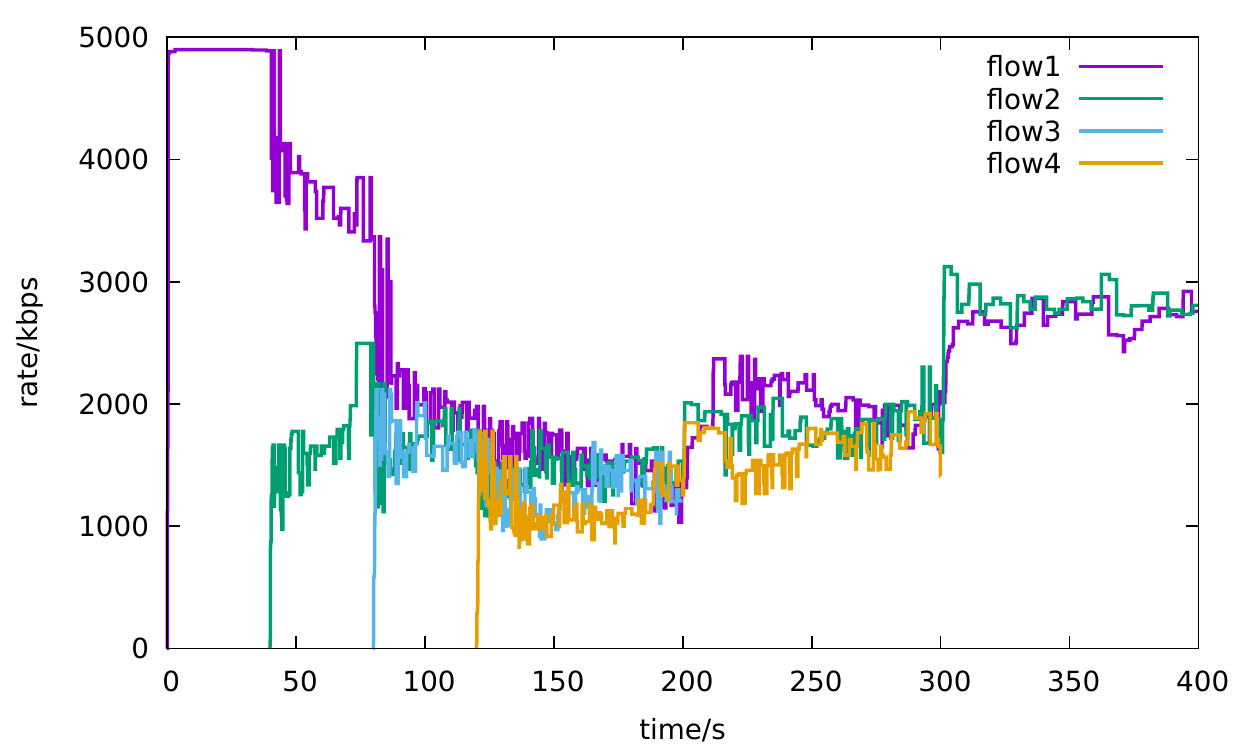}
\end{minipage}}
\subfigure[C3]{
\begin{minipage}[t]{0.5\linewidth}
    \includegraphics[width = 1\linewidth]{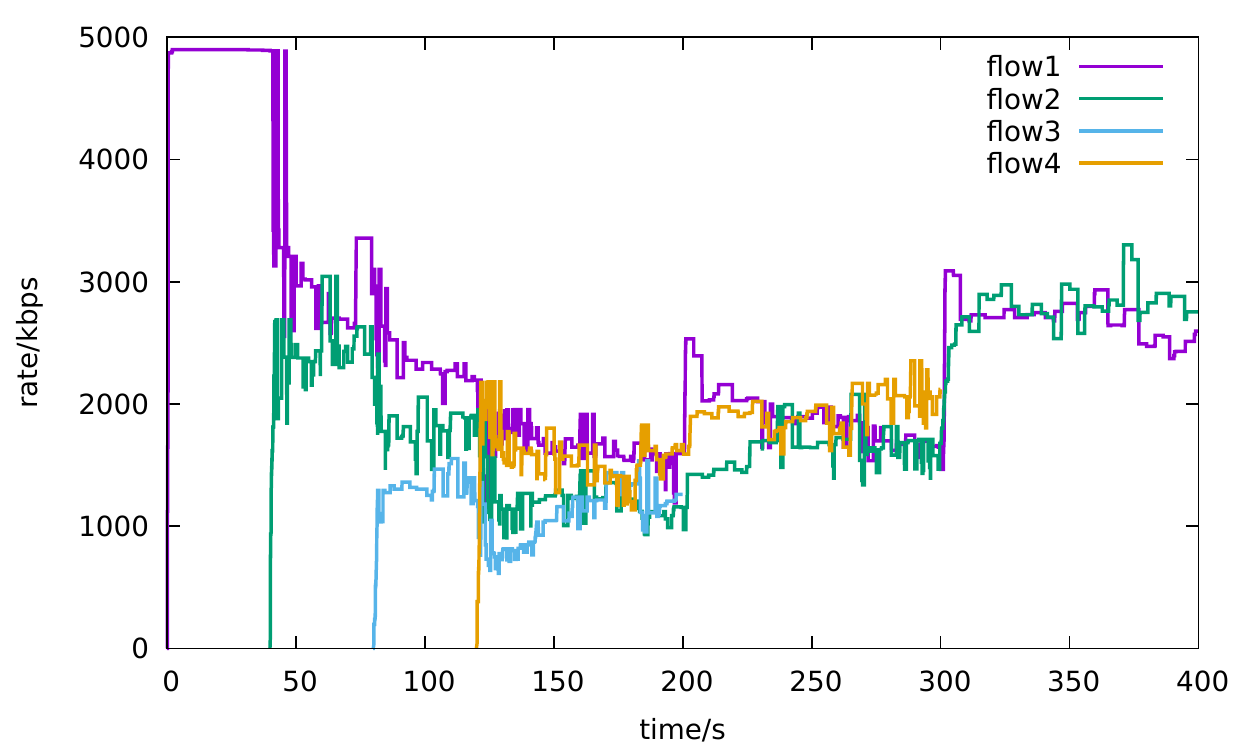}  
\end{minipage}}
\end{tabular}
\caption{Rate dynamics of BBR v2 flows}
\label{Fig:bbrv2-1-3} 
\end{figure}
\subsection{RTT unfairness}
As indicated in several reports, BBR flows traversing the bottleneck link with same RTT converge to a fair share bandwidth line within their group, but these flows with longer RTT can achieve higher throughput, contrary to the classic Reno algorithm. Reno algorithm favors towards shorter RTT flows. The Reno flows with shorter RTT can get acknowledge packet more quickly and the congestion window can be increased faster in these flows. Such RTT unfairness property in BBR can be easily manipulated by malicious receiver for data transmission acceleration by delaying the acknowledge packets.

The dumbbell topology is applied to reproduce the RTT unfairness phenomenon of BBR algorithm. The bottleneck link is $l2$. The capacities of $l1, l3, l4, l5$ are 100 Mbps. The propagation delay values for $l1, l3$ are 10 milliseconds. 20 milliseconds is configured for $l4, l5$. Nine experiments cases are tested. The parameters to configure $l2$ are in each case given in Table \ref{tab:l2rtt}. Two flows are applied. $flow1 $starts from $n0$ to destination $n4$ (path1) and $flow2$ starts from $n1$ to destination $n5$ (path2). The max round trip propagation delay of path1 and path2 is denoted as $RPT_{max}$. $Qdelay$ is $1.5* RPT_{max}$. In other word, the buffer length of bottleneck link is $1.5*BDP$. The buffer length of Other links is $BW* Qdelay$.

The running time of each simulation lasts 200 seconds. Both flows are running in the whole simulation time. The average throughout is calculated as Equation \eqref{eq:rate}. $bytes$ is the length of all received packets. The jain’s fairness index \cite{Jain1984quantitative} is exploited to indicate how fair the bandwidth is shared when flows competing for bandwidth resource. The way to compute the jain’s fairness index is shown in \eqref{eq:jain}. The closer Jain's fairness index is to 1, the better in terms of bandwidth allocation fairness.

In this part, only two flows are involved in each case. As we found in some case, Jain's fairness index may not work well to reflect the throughput variance and the throughput ratio of the two flow is computed as Equation \eqref{eq:ratio}. The final results are shown in Table \ref{tab:rttfair}. $\overline x_1$ and $\overline x_2$ are the average throughput (in unit of kbps) of $flow1$ and $flow2$ in respectively.
\begin{equation}
\label{eq:rate}
\overline x=\frac{bytes}{duration}
\end{equation}
\begin{equation}
\label{eq:jain}
J=\frac{(\sum_i^{n}{x_i})^2}{n*(\sum_i^{n}{x_i^2})}
\end{equation}
\begin{equation}
\label{eq:ratio}
r=\frac{\overline x_{max}}{\overline x_{min}}
\end{equation}

Some conclusions can be made based on Table \ref{tab:rttfair}. BBR $flow2$ with longer RTT acquires more than four times the throughput of $flow1$ in Case7. As the RTT ratio gets smaller in different cases, the rates of two flows are get closer and the jain’s fairness is also increase. The reason for RTT unfairness is related to buffer occupation in intermediate routers. The flow with large RTT will send more packets out when the bottleneck is already in congestion. It will get larger bandwidth estimation than the flow with shorter RTT flow.

In BBR’ and BBRPlus cases, the throughput ratios are not so large as in BBR cases. The RTT unfairness issue has been significantly improved. Both algorithms will drain the inflight packet to match the estimated $BDP$ in the probe down phase and lower queue delay can be achieved as shown in Figure \ref{Fig:owd-comapre}. The action to drain to the target will give other flows an opportunity to probe more available bandwidth. That’s the reason behind such improvement.

The RTT unfairness is not so severe in BBR v2 test cases as BBR. The results on Reno verify the conclusion that the flow with shorter RTT can gain higher throughout.
\begin{table}[]
\centering
\caption{the configuration of $l2$ to test RTT unfairness}
\label{tab:l2rtt}
\begin{tabular}{|c|c|c|c|}
\hline
Case & Bandwidth& Propagation delay& Q(bw*Qdelay)\\ \hline
1 & 4Mbps & 10ms & 4Mbps*150ms \\ \hline
2 & 4Mbps & 20ms & 4Mbps*180ms \\ \hline
3 & 4Mbps & 30ms & 4Mbps*210ms \\ \hline
4 & 6Mbps & 10ms & 6Mbps*150ms \\ \hline
5 & 6Mbps & 20ms & 6Mbps*180ms \\ \hline
6 & 6Mbps & 30ms & 6Mbps*210ms \\ \hline
7 & 8Mbps & 10ms & 8Mbps*150ms \\ \hline
8 & 8Mbps & 20ms & 8Mbps*180ms \\ \hline
9 & 8Mbps & 30ms&  8Mbps*210ms \\ \hline
\end{tabular}
\end{table} 
\begin{table*}[]
\centering
\caption{The calculated results in RTT unfairness simulation}
\label{tab:rttfair}
\resizebox{\textwidth}{15mm}{
\begin{tabular}{|c|c|c|c|c|c|c|c|c|c|}
\hline
\multirow{2}{*}{algo} & 1    & 2   & 3   & 4   & 5   & 6   & 7   & 8   & 9   \\ \cline{2-10} 
                      & \multicolumn{9}{c|}{($\overline{x_1}$, $\overline{x_2}$, jain's fairness index, r)} \\ \hline
BBR                   &840, 2970, 0.76, 3.53&1029, 2767, 0.83, 2.69&1177, 2606, 0.88, 2.21&1154, 4543, 0.74, 3.94&1511, 4166, 0.82, 2.76&1703, 3947, 0.86, 2.32&1421, 6164, 0.72, 4.34&1944, 5611, 0.81, 2.89&2259, 5269, 0.86, 2.33  \\ \hline
BBR'                  &1668, 2159, 0.98, 1.29&1712, 2098, 0.99, 1.23&1800, 2001, 1.0, 1.11&2498, 3221, 0.98, 1.29&2508, 3168, 0.99, 1.26&2729, 2964, 1.0, 1.09&3329, 4273, 0.98, 1.28&3344, 4216, 0.99, 1.26&3517, 4055, 0.99, 1.15  \\ \hline
BBRPlus               &1713, 2105, 0.99, 1.23&1785, 2017, 1.0, 1.13&1774, 2021, 1.0, 1.14&2594, 3126, 0.99, 1.21&2580, 3098, 0.99, 1.2&2656, 3024, 1.0, 1.14&3381, 4223, 0.99, 1.25&3394, 4164, 0.99, 1.23&3389, 4182, 0.99, 1.23   \\ \hline
Tsunami               &902, 2914, 0.78, 3.23&1103, 2706, 0.85, 2.45&1336, 2462, 0.92, 1.84&1226, 4486, 0.75, 3.66&1621, 4080, 0.84, 2.52&1792, 3873, 0.88, 2.16&1418, 6182, 0.72, 4.36&1977, 5601, 0.81, 2.83&2360, 5181, 0.88, 2.19     \\ \hline
BBR+                  &1345, 2481, 0.92, 1.84&1500, 2305, 0.96, 1.54&1682, 2116, 0.99, 1.26&2073, 3653, 0.93, 1.76&2235, 3436, 0.96, 1.54&2495, 3181, 0.99, 1.27&2543, 5054, 0.9, 1.99&2999, 4556, 0.96, 1.52&3334, 4225, 0.99, 1.27  \\ \hline
BBRv2                 &1695, 2182, 0.98, 1.29&1663, 2212, 0.98, 1.33&2025, 1850, 1.0, 1.09&2474, 3343, 0.98, 1.35&2337, 3469, 0.96, 1.48&3338, 2473, 0.98, 1.35&3649, 4102, 1.0, 1.12&4134, 3602, 1.0, 1.15&4465, 3279, 0.98, 1.36 \\ \hline
Cubic                 &1716, 2166, 0.99, 1.26&1625, 2257, 0.97, 1.39&2429, 1452, 0.94, 1.67&2845, 2978, 1.0, 1.05&3239, 2583, 0.99, 1.25&3501, 2320, 0.96, 1.51&3501, 4263, 0.99, 1.22&3524, 4239, 0.99, 1.2&4824, 2936, 0.94, 1.64   \\ \hline
Reno                  &2237, 1645, 0.98, 1.36&2062, 1820, 1.0, 1.13&2239, 1642, 0.98, 1.36&3406, 2417, 0.97, 1.41&3393, 2429, 0.97, 1.4&3370, 2451, 0.98, 1.37&4219, 3546, 0.99, 1.19&4154, 3608, 1.0, 1.15&4331, 3429, 0.99, 1.26 \\ \hline
\end{tabular}}
\end{table*}
\subsection{Channel utilization}
The channel utilization of these congestion control algorithms is tested in links with random loss. The configuration of the point to point channel ($n2$ to $n3$) remains unchanged as in Table \ref{tab:l2}. Only five cases (C2, C5, C7, C9, C10) are involved and the configured buffer length of the bottleneck is $1.5*BDP$. In each case, the random packet loss rates are 1\%, 3\% and 5\%. The four flows are running in the whole simulation process.

The channel utilization of all flows is calculates as Equation \eqref{eq:util}. $bytes_i$ is the length of all received packets at application layer of flow $i$. $cap$ is the bandwidth of the bottleneck link and $duration$ is the simulation running time. The final results are given in Table \ref{tab:lossutil}.

When no random packet loss is existence, these algorithms achieve channel bandwidth utilization above 90\%. BBR v2 can achieve channel utilization about 97\%, similar to the two buffer filling algorithms Reno and Cubic. The link utilization of the three algorithms (BBR, BBR’ and BBRPlus) is less affected by random packet loss. With 5\% random packet loss rate, the channel utilization of BBR v2 flows is quite low in C7 and C9. 
\begin{equation}
\label{eq:util}
util=\frac{\sum_{i}{bytes_i}}{cap*duration}
\end{equation}
%\begin{table}[]
%\caption{channel utilization without random loss}
%\label{tab:util}
%\scalebox{0.8}{
%\begin{tabular}{|c|c|c|c|c|c|c|c|c|}
%\hline
%Case & BBR & BBR' & BBRPlus & BBR+ & Tsu & BBRv2 & Reno & Cubic \\ \hline
%1    &0.95&0.95&0.95&0.95&0.95&0.97&0.97&0.97\\ \hline
%2    &0.95&0.95&0.95&0.95&0.95&0.97&0.97&0.97\\ \hline
%3    &0.95&0.95&0.95&0.95&0.95&0.97&0.97&0.97\\ \hline
%4    &0.95&0.95&0.95&0.95&0.95&0.97&0.97&0.97\\ \hline
%5    &0.95&0.95&0.95&0.95&0.95&0.97&0.97&0.97\\ \hline
%6    &0.95&0.95&0.95&0.95&0.95&0.97&0.97&0.97\\ \hline
%7    &0.94&0.94&0.95&0.94&0.94&0.97&0.96&0.96\\ \hline
%8    &0.95&0.95&0.95&0.95&0.95&0.97&0.97&0.97\\ \hline
%9    &0.94&0.94&0.94&0.94&0.94&0.97&0.96&0.96\\ \hline
%10   &0.95&0.95&0.95&0.95&0.95&0.97&0.97&0.97\\ \hline
%11   &0.95&0.95&0.95&0.95&0.95&0.97&0.97&0.97\\ \hline
%\end{tabular}}
%\end{table}
\begin{table}[]
\caption{channel utilization with random loss}
\label{tab:lossutil}
\begin{tabular}{|c|c|c|c|c|c|c|}
\hline
algo                     & rand loss(\%)& C2 & C5 & C7 & C9 & C10 \\ \hline
\multirow{4}{*}{BBR}     & 0    &0.95&0.95 &0.94&0.94&0.95 \\ \cline{2-7} 
                         & 1    &0.95&0.94 &0.93&0.92&0.94 \\ \cline{2-7} 
                         & 3    &0.92&0.92 &0.91&0.90&0.92 \\ \cline{2-7} 
                         & 5    &0.91&0.90 &0.89&0.89&0.90 \\ \hline
\multirow{4}{*}{BBR'}    & 0    &0.95&0.95 &0.94&0.94&0.95 \\ \cline{2-7} 
                         & 1    &0.94&0.94 &0.93&0.93&0.94 \\ \cline{2-7} 
                         & 3    &0.92&0.92 &0.91&0.91&0.92 \\ \cline{2-7} 
                         & 5    &0.90&0.90 &0.89&0.89&0.90 \\ \hline
\multirow{4}{*}{BBRPlus} & 0    &0.95&0.95 &0.95&0.94&0.95 \\ \cline{2-7} 
                         & 1    &0.94&0.94 &0.93&0.92&0.94 \\ \cline{2-7} 
                         & 3    &0.92&0.92 &0.91&0.90&0.92 \\ \cline{2-7} 
                         & 5    &0.90&0.90 &0.89&0.88&0.90 \\ \hline
\multirow{4}{*}{BBR v2}  & 0    &0.97&0.97 &0.97&0.97&0.97 \\ \cline{2-7} 
                         & 1    &0.96&0.96 &0.95&0.96&0.96 \\ \cline{2-7} 
                         & 3    &0.94&0.93 &0.87&0.84&0.92 \\ \cline{2-7} 
                         & 5    &0.91&0.91 &0.62&0.51&0.81 \\ \hline
\multirow{4}{*}{Cubic}   & 0    &0.97&0.97 &0.96&0.96&0.97 \\ \cline{2-7} 
                         & 1    &0.96&0.96 &0.70&0.62&0.92 \\ \cline{2-7} 
                         & 3    &0.93&0.86 &0.38&0.33&0.53 \\ \cline{2-7} 
                         & 5    &0.76&0.64 &0.28&0.25&0.40 \\ \hline
\multirow{4}{*}{Reno}    & 0    &0.97&0.97 &0.96&0.96&0.97 \\ \cline{2-7} 
                         & 1    &0.96&0.96 &0.85&0.75&0.96 \\ \cline{2-7} 
                         & 3    &0.94&0.92 &0.46&0.40&0.64 \\ \cline{2-7} 
                         & 5    &0.87&0.78 &0.34&0.30&0.48 \\ \hline
\end{tabular}
\end{table}
\subsection{Responsiveness}
In cellular access network or wireless network, channel throughput can present drastic change in a short time span due to noise interference and fading. The performance of BBR is tested in cellular network in \cite{Atxutegi2018Use}. Here, the point to point link $n2$ to $n3$ is used to test whether these algorithm can make fast response to link bandwidth change. The link capacity is changed every 50 seconds from 1Mbps to 4Mbps to simulate link throughout change. The propagation delay is 50 milliseconds. Two flows are involved and the simulation process lasts 400 second. 

\begin{figure*}[!htb]
\begin{tabular}{cc}
\subfigure[BBR]{
\begin{minipage}[t]{0.3\linewidth}
    \includegraphics[width = 1\linewidth]{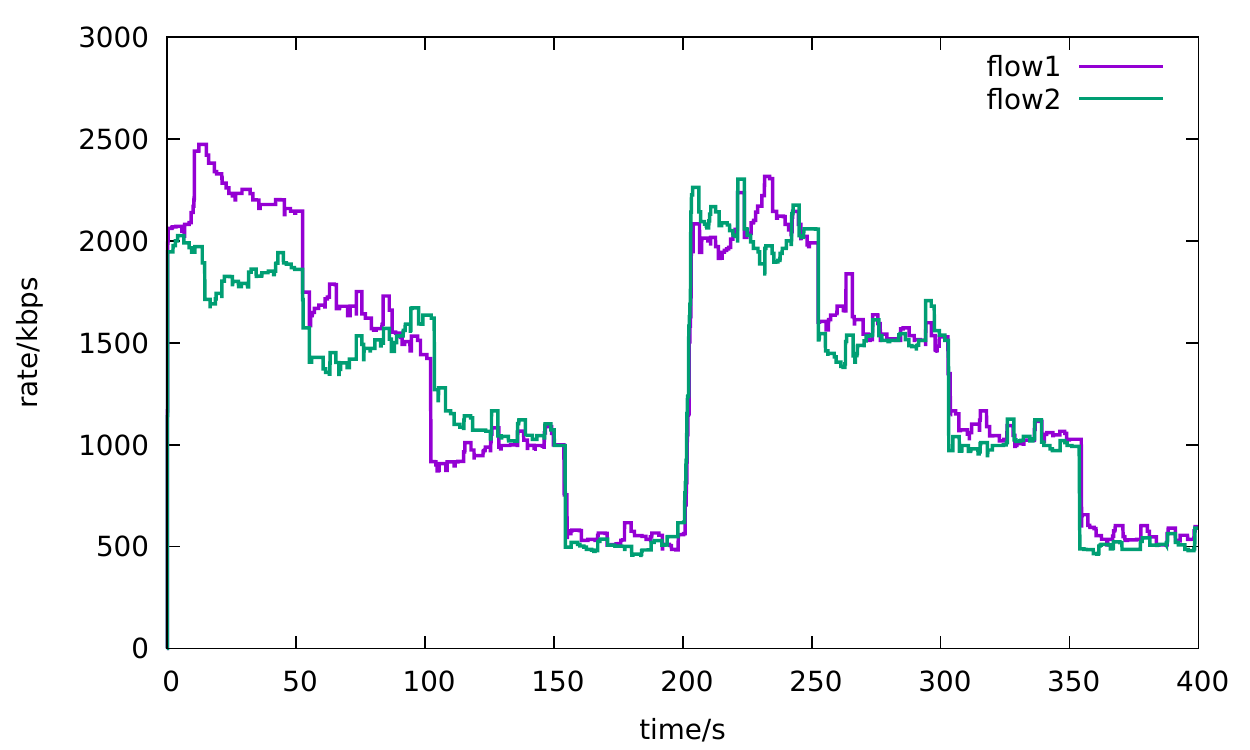}
\end{minipage}}
\subfigure[BBR']{
\begin{minipage}[t]{0.3\linewidth}
    \includegraphics[width = 1\linewidth]{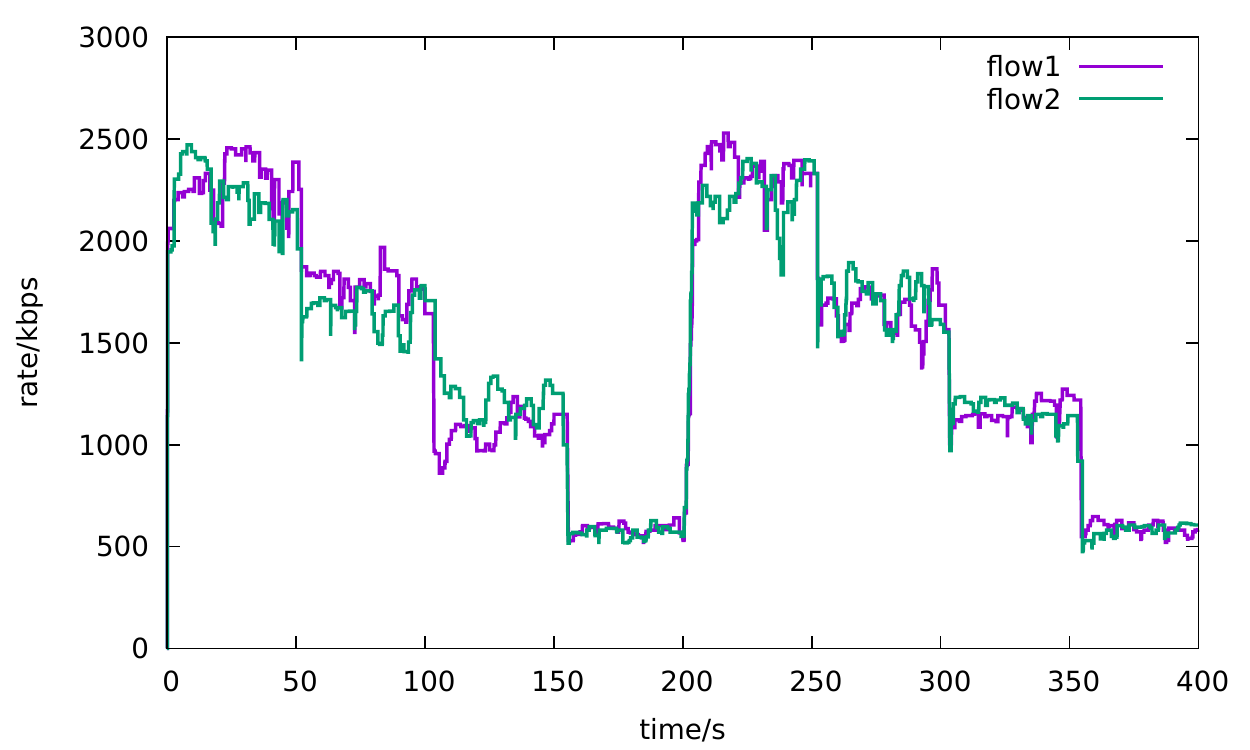}  
\end{minipage}}
\subfigure[BBRPlus]{
\begin{minipage}[t]{0.3\linewidth}
    \includegraphics[width = 1\linewidth]{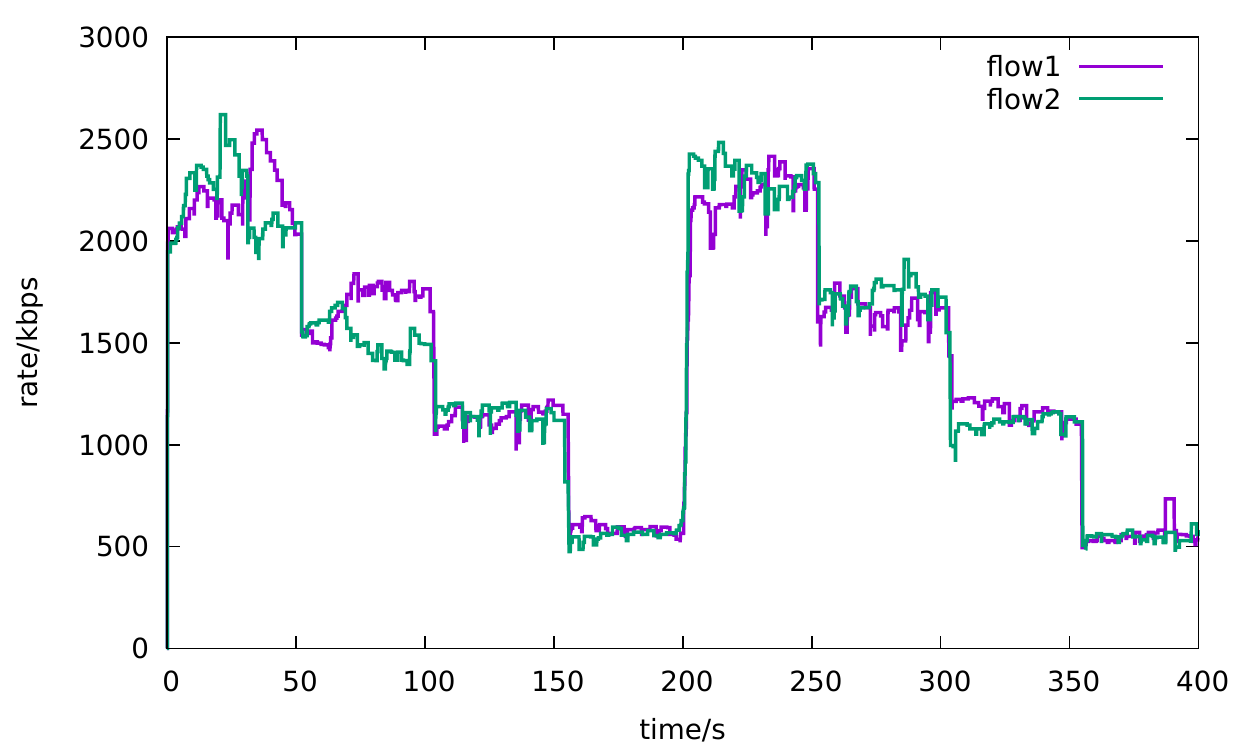}  
\end{minipage}}
\\
\subfigure[BBR+]{
\begin{minipage}[t]{0.3\linewidth}
    \includegraphics[width = 1\linewidth]{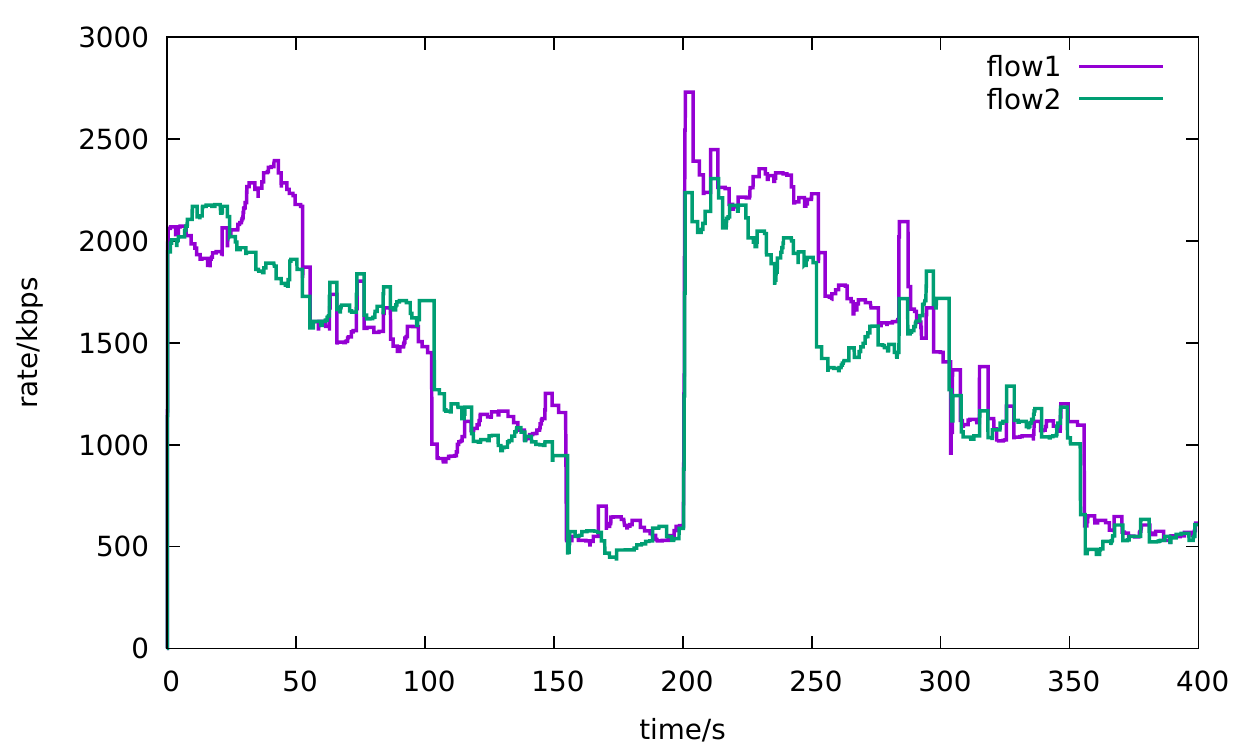}  
\end{minipage}}
\subfigure[Tsunami]{
\begin{minipage}[t]{0.3\linewidth}
    \includegraphics[width = 1\linewidth]{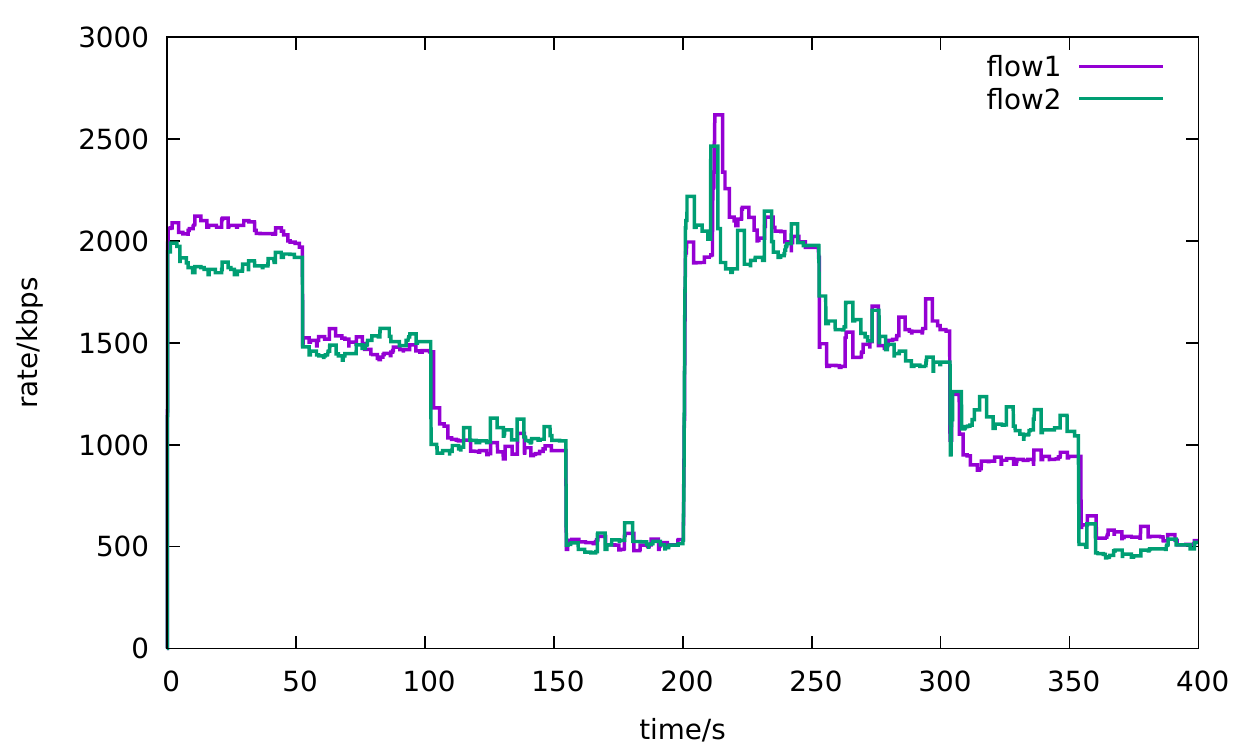}  
\end{minipage}}
\subfigure[BBRv2]{
\begin{minipage}[t]{0.3\linewidth}
    \includegraphics[width = 1\linewidth]{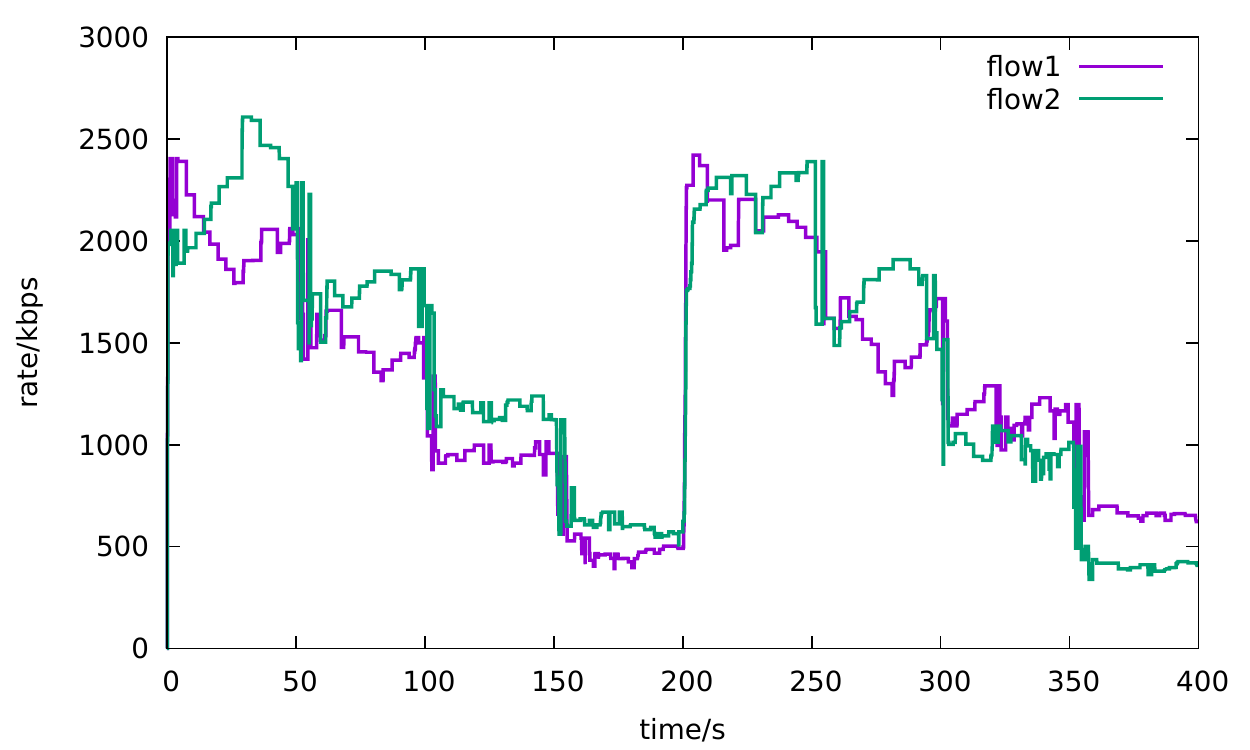}  
\end{minipage}}
\end{tabular}
\caption{Rate dynamic of flows in link with variable capacity}
\label{Fig:resp} 
\end{figure*}

The rate adjustment process of each flow is shown in Figure \ref{Fig:resp}. All these algorithms can adapt well the throughput of the flow as the link rate changes. The average packet loss rate, average packet transmission delay and channel utilization are calculated in Table \ref{tab:resp}. Since the buffer length is configured as $1.5*4Mbps*100ms$, the loss rate of flows with BBR like algorithms is quite small (below 1\%). The packet transmission delay values are higher in Tsunami, BBR+ and BBRv2. BBR’ can achieve the lowest packet transmission delay and the second is BBRPlus.
\begin{table}[]
\caption{The statistical results of the responsive experiments}
\label{tab:resp}
\begin{tabular}{|c|c|c|c|}
\hline
        & loss & $\overline{owd}$ & utility \\ \hline
BBR     &0.001&170.21&0.95\\ \hline
BBR'    &0.001&117.35&0.95\\ \hline
BBRPlus &0.001&139.15&0.95\\ \hline
Tsunami &0.003&182.37&0.95\\ \hline
BBR+    &0.005&186.19&0.95\\ \hline
BBrv2   &0.006&224.58&0.97\\ \hline
Cubic   &0.009&272.23&0.97\\ \hline
Reno    &0.020&257.26&0.97\\ \hline
\end{tabular}
\end{table}
\begin{table}[]
\centering
\caption{the configuration of $l2$ to test inter protocol fairness}
\label{tab:l2inter}
\begin{tabular}{|c|c|c|c|}
\hline
Case & Bandwidth& Propagation delay& Queue length\\ \hline
1 & 4Mbps & 50ms & 4Mbps*100ms \\ \hline
2 & 4Mbps & 50ms & 4Mbps*150ms \\ \hline
3 & 4Mbps & 50ms & 4Mbps*200ms \\ \hline
4 & 6Mbps & 50ms & 6Mbps*100ms \\ \hline
5 & 6Mbps & 50ms & 6Mbps*150ms \\ \hline
6 & 6Mbps & 50ms & 6Mbps*200ms \\ \hline
7 & 8Mbps & 50ms & 8Mbps*150ms \\ \hline
8 & 10Mbps & 50ms & 10Mbps*150ms \\ \hline
9 & 12Mbps & 50ms&  12Mbps*150ms \\ \hline
\end{tabular}
\end{table}
\begin{table}[]
\caption{Results of Jain's fairness index and ratio}
\label{tab:inter}
\resizebox{\linewidth}{10mm}{
\begin{tabular}{|c|c|c|c|c|c|c|c|c|c|}
\hline
        & 1   & 2   & 3   & 4   & 5   & 6   & 7   & 8   & 9   \\ \hline
        & \multicolumn{9}{c|}{(Jain's fairness index, ratio)} \\ \hline
BBR     &0.55, 13.61&0.95, 1.7&1.0, 1.09&0.5, 17.08&0.97, 1.52&0.98, 1.42&0.96, 1.63&0.97, 1.53&0.97, 1.59  \\ \hline
BBR'    &0.93, 1.94&0.97, 1.68&0.99, 1.31&0.99, 1.38&0.98, 1.48&0.97, 1.57&0.96, 1.73&0.95, 1.86&0.95, 1.81  \\ \hline
BBRPlus &0.98, 1.48&0.98, 1.4&1.0, 1.13&0.99, 1.22&1.0, 1.17&0.99, 1.28&0.99, 1.28&0.98, 1.51&0.98, 1.35  \\ \hline
BBR+    &0.65, 5.87&0.81, 3.0&0.95, 1.69&0.54, 7.93&0.92, 1.97&0.97, 1.58&0.86, 2.46&0.96, 1.6&0.92, 2.0  \\ \hline
Tsunami &0.4, 26.08&0.91, 2.04&0.74, 4.03&0.38, 34.44&0.91, 2.13&0.74, 3.67&0.86, 2.62&0.86, 2.56&0.95, 1.76  \\ \hline
BBRv2   &0.96, 1.65&0.98, 1.46&0.99, 1.35&0.99, 1.31&0.97, 1.44&1.0, 1.13&1.0, 1.13&0.97, 1.61&0.95, 1.73  \\ \hline
\end{tabular}}
\end{table}
\begin{figure*}[!htb]
\begin{tabular}{cc}
\subfigure[BBR vs Cubic]{
\begin{minipage}[t]{0.3\linewidth}
    \includegraphics[width = 1\linewidth]{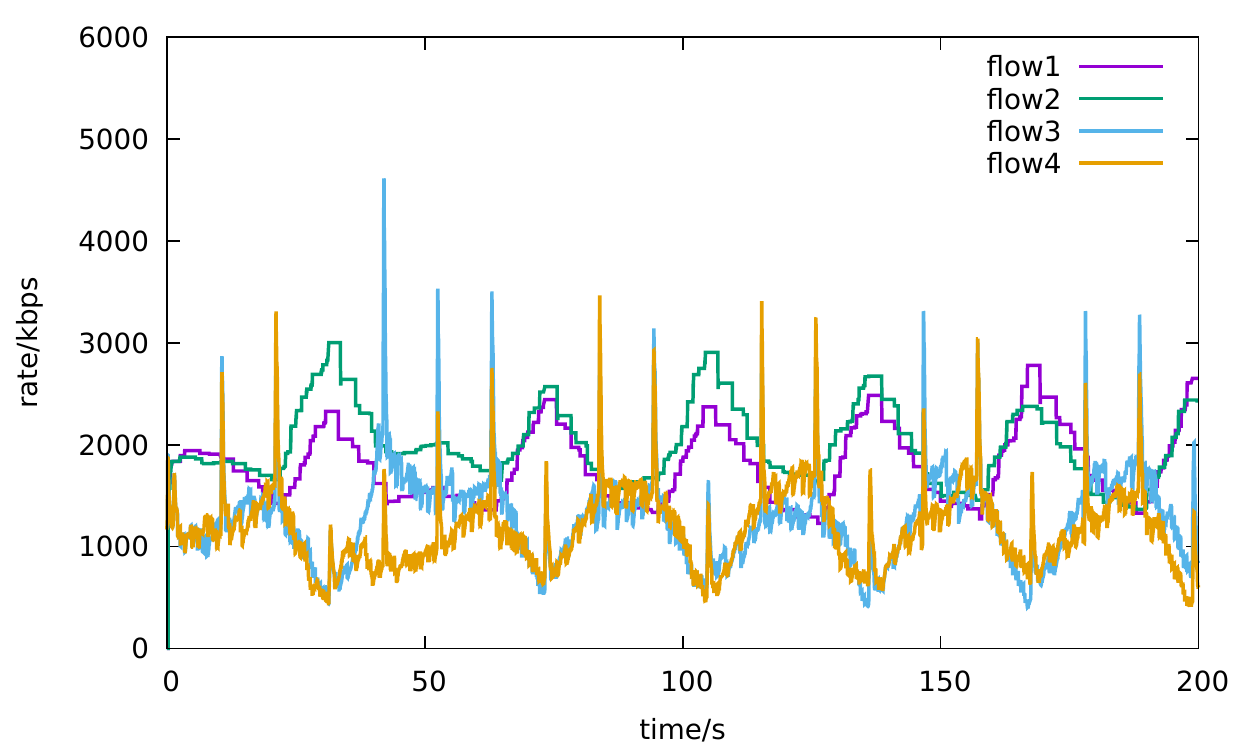}
\end{minipage}}
\subfigure[BBR' vs Cubic]{
\begin{minipage}[t]{0.3\linewidth}
    \includegraphics[width = 1\linewidth]{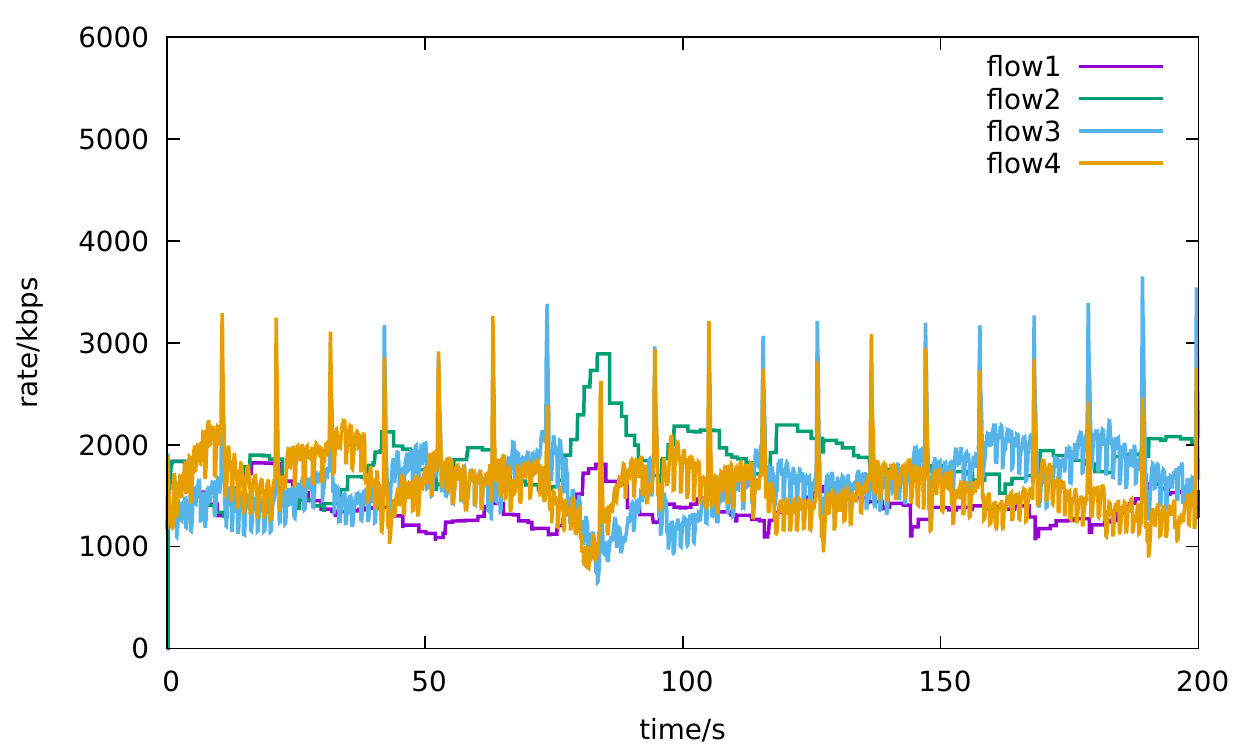}  
\end{minipage}}
\subfigure[BBRPlus vs Cubic]{
\begin{minipage}[t]{0.3\linewidth}
    \includegraphics[width = 1\linewidth]{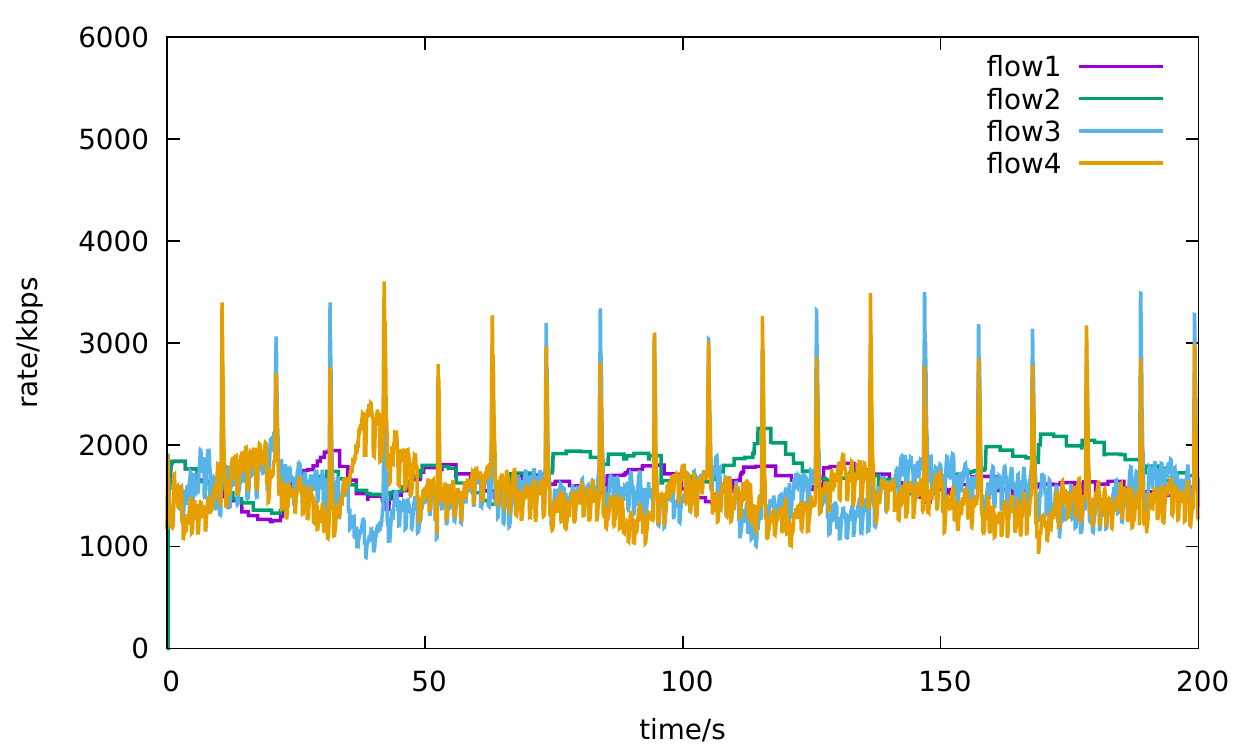}  
\end{minipage}}
\\
\subfigure[BBR+ vs Cubic]{
\begin{minipage}[t]{0.3\linewidth}
    \includegraphics[width = 1\linewidth]{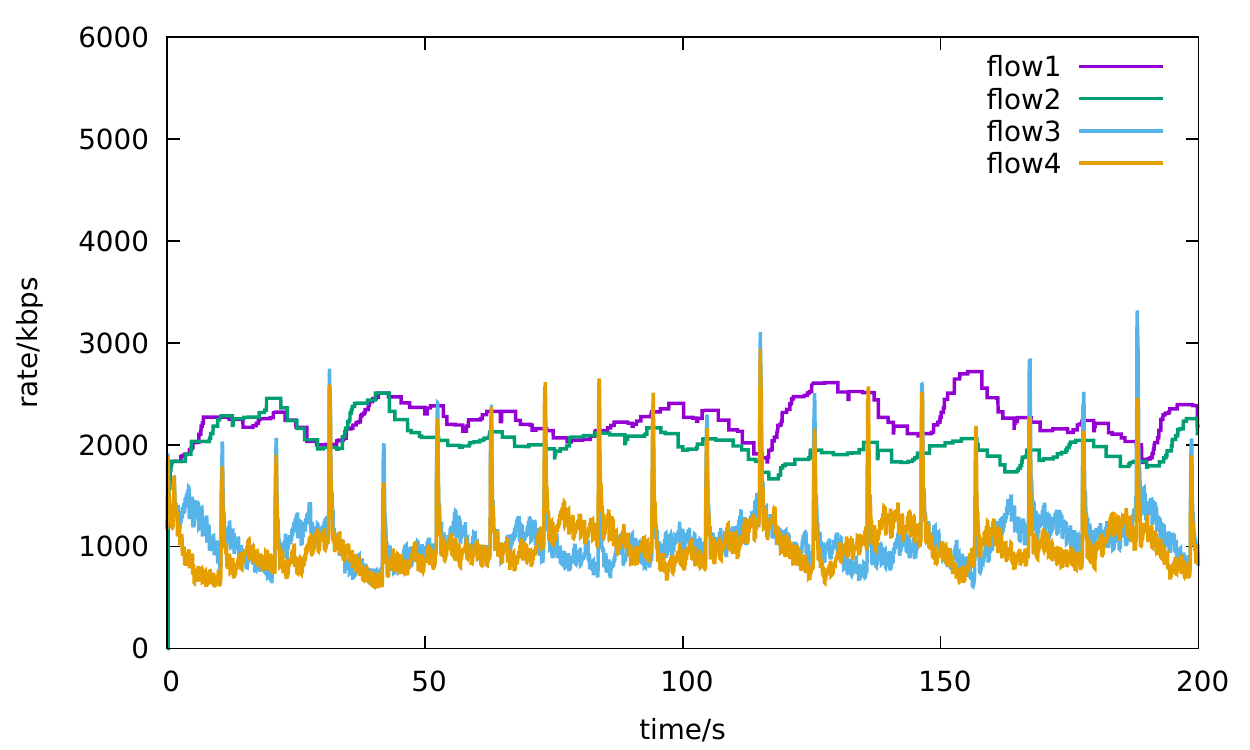}  
\end{minipage}}
\subfigure[Tsunami vs Cubic]{
\begin{minipage}[t]{0.3\linewidth}
    \includegraphics[width = 1\linewidth]{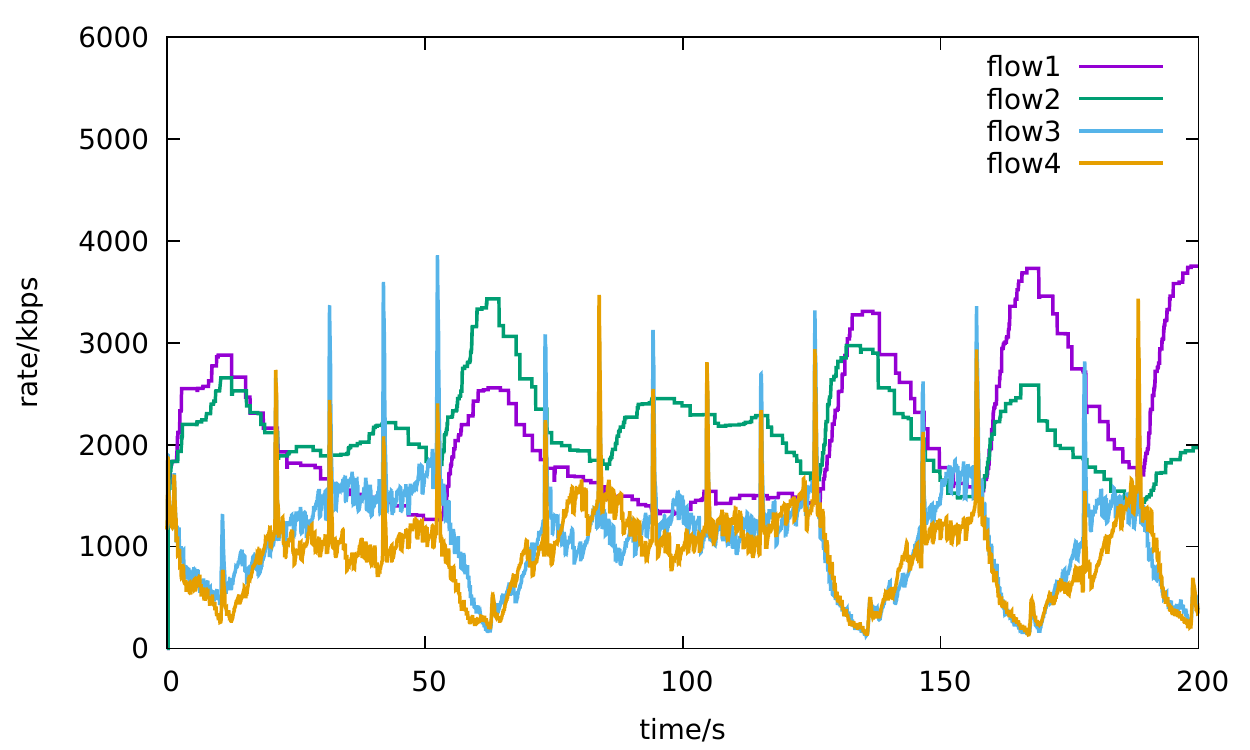}  
\end{minipage}}
\subfigure[BBRv2 vs Cubic]{
\begin{minipage}[t]{0.3\linewidth}
    \includegraphics[width = 1\linewidth]{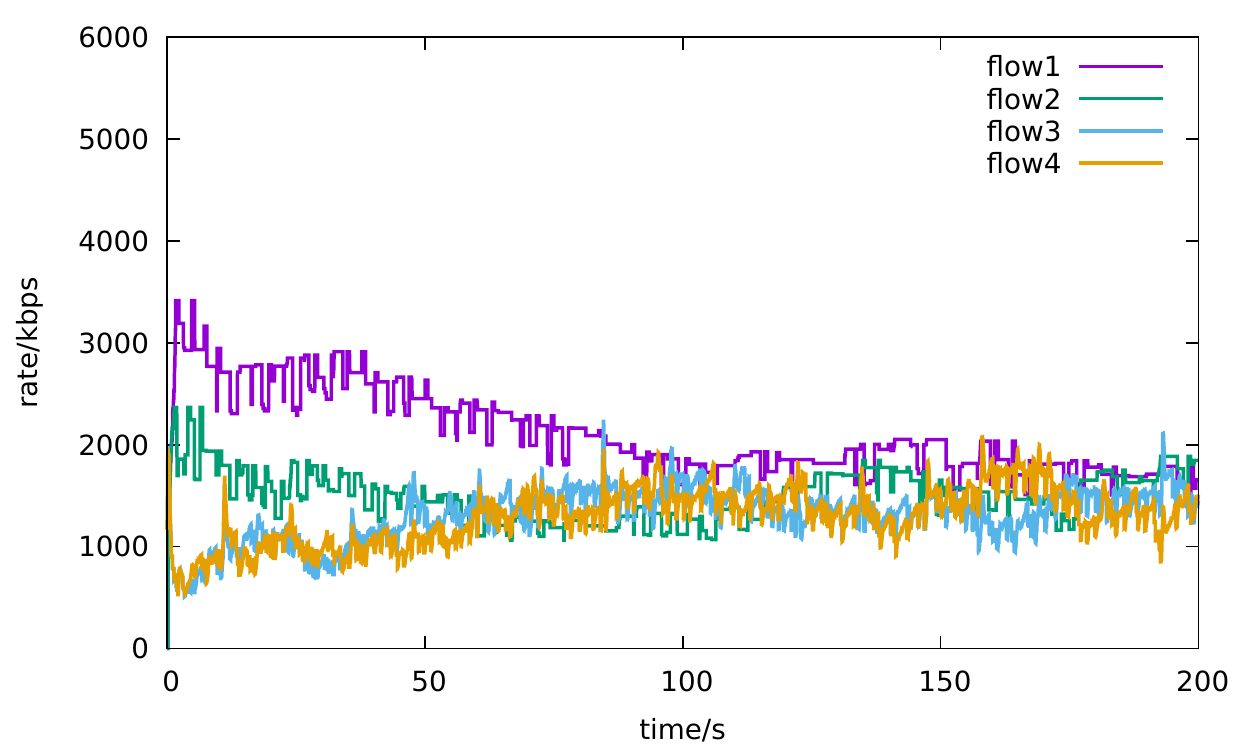}  
\end{minipage}}
\end{tabular}
\caption{Rate dynamic of flows with different congestion control algorithms}
\label{Fig:inter} 
\end{figure*}
\subsection{Inter Protocol Competition}
In real networks, a bottleneck link is multiplexed by multiple flows following different congestion control algorithm. And one of the goals of BBR v2 is to make better coexistence with Cubic and Reno flows. The performance of these algorithms are evaluated when the Cubic flows are presence. Total nine experiments are designed and the configuration of $l2$ is shown in Table \ref{tab:l2inter}. The simulation process lasts 200 seconds. In each case, the congestion control algorithm used by flow3 and flow4 is Cubic.

The jain’s fairness and the throughout ratio between the flow with the maximum rate and the flow with the minimum rate are calculated in Table \ref{tab:inter}. As example, the rate dynamic over time of the four flows in Case 5 is plotted in Figure \ref{Fig:inter}.

In Case1 and Case4, the buffer length is equal to $BDP$. There are high packet loss rate when BBR, BBR+ and Tsunami are applied for flow1 and flow2 as congestion control algorithm. The two Cubic flows suffer great loss in throughout due to these packet loss event. Most of bandwidth is occupied by flow1 and flow2. The jain’s fairness index is quite low and the throughput ratio is high in this two test cases. As the link buffer increase (Case2, Case3, Case5 and Case6), Cubic flows can achieve higher throughput, but the throughputs of flow1 and flow2 still dominate. This conclusion gets support in Figure \ref{Fig:inter}(a), Figure \ref{Fig:inter}(d), Figure \ref{Fig:inter}(e). The rate of BBR flows (Figure \ref{Fig:inter}(a)), Tsunami flows (Figure \ref{Fig:inter}(e)) shows oscillation when competing bandwidth with Cubic flows.

When BBRPlus flows sharing bottleneck link with Cubic flows, the value of Jain’s fairness is quite close to 1, and the rates of BBRPlus flows and Cubic flows are quite close in Case5 as shown in Figure \ref{Fig:inter}(c). In Figure \ref{Fig:inter}(f), the rates of BBRv2 flows and Cubic flows converge close to the fairness line. Both BBRPlus and BBRv2 can be friendly to Cubic.
\section{Discussion and Conclusion}
We evaluate the performance of BBR, BBR v2 and other algorithms (BBR’, BBRPlus, BBR+, Tsunami) modified from BBR on ns3 platform. Bandwidth allocation fairness, packet loss rate, link queue buffer occupation, RTT unfairness, channel utilization, rate responsiveness in variable links and inter protocol fairness are evaluated or measured in this article.

In link with shallow buffer, the flows taking BBR, BBR+ and Tsunami as rate control can not converge the fairness bandwidth line and high packet loss rate is introduced. In the probe down down, the inflight packet will be drained to match the estimated bandwidth delay product in BBR’ and BBRPlus. These two can achieve lower transmission delay. BBR’ and BBRPlus can maintain well bandwidth fairness property even in shallow buffer link. But the rates of BBR’ flows have larger range of variation compared with BBR. Tsunami flow will send packet with aggressive rate higher than its estimated bandwidth and introduces high transmission delay and highest packet loss rate. Tsunami will lead severe congestion, and such egoistic implementation is not recommended to applied in real networks.

As for the RTT unfairness issue, the results in BBR’, BBRPlus and BBR v2 show improvement when compared with BBR test cases. The average packet transmission delay values in BBR v2 flows are lower than these delay values in Cubic/Reno flows.

All tested algorithms can get high channel utilization When bottleneck has no random packet loss. With 5\% random loss rate, BBR v2 flows can get low channel utilization in two test cases.

BBR v2 indeed achieve its goal to make better coexistence with Cubic/Reno flows. But its claim to achieve lower queue delay is not obvious when compared with BBR. The inter protocol friendliness can also be achieved in BBRPlus to some extent. 

BBRPlus is an good improvement to BBR and is highly recommended to be applied in this paper.

\bibliographystyle{elsarticle-num}
\bibliography{bbr-simu,ldc}

\begin{thebibliography}{10}
\expandafter\ifx\csname url\endcsname\relax
  \def\url#1{\texttt{#1}}\fi
\expandafter\ifx\csname urlprefix\endcsname\relax\def\urlprefix{URL }\fi
\expandafter\ifx\csname href\endcsname\relax
  \def\href#1#2{#2} \def\path#1{#1}\fi

\bibitem{Jacobson1988Congestion}
V.~Jacobson, Congestion avoidance and control, in: ACM SIGCOMM computer
  communication review, Vol.~18, ACM, 1988, pp. 314--329.

\bibitem{Lin2019Extensive}
J.~Lin, L.~Cui, Y.~Zhang, F.~P. Tso, Q.~Guan,
  \href{http://www.sciencedirect.com/science/article/pii/S1389128618311265}{Extensive
  evaluation on the performance and behaviour of tcp congestion control
  protocols under varied network scenarios}, Computer Networks 163 (2019)
  106872.
\newblock \href
  {http://dx.doi.org/https://doi.org/10.1016/j.comnet.2019.106872}
  {\path{doi:https://doi.org/10.1016/j.comnet.2019.106872}}.
\newline\urlprefix\url{http://www.sciencedirect.com/science/article/pii/S1389128618311265}

\bibitem{Turkovic2019Fifty}
B.~Turkovic, F.~A. Kuipers, S.~Uhlig, Fifty shades of congestion control: A
  performance and interactions evaluation, arXiv preprint arXiv:1903.03852.

\bibitem{Gettys2011Bufferbloat}
J.~Gettys, K.~Nichols, Bufferbloat: Dark buffers in the internet, Queue 9~(11)
  (2011) 1--15.

\bibitem{Cardwell2016BBR}
N.~Cardwell, Y.~Cheng, C.~S. Gunn, S.~H. Yeganeh, V.~Jacobson, Bbr:
  Congestion-based congestion control, Queue 14~(5) (2016) 50:20--50:53.
\newblock \href {http://dx.doi.org/10.1145/3012426.3022184}
  {\path{doi:10.1145/3012426.3022184}}.

\bibitem{Hock2017Experimental}
M.~{Hock}, R.~{Bless}, M.~{Zitterbart}, Experimental evaluation of bbr
  congestion control, in: 2017 IEEE 25th International Conference on Network
  Protocols (ICNP), 2017, pp. 1--10.
\newblock \href {http://dx.doi.org/10.1109/ICNP.2017.8117540}
  {\path{doi:10.1109/ICNP.2017.8117540}}.

\bibitem{Scholz2018Towards}
D.~{Scholz}, B.~{Jaeger}, L.~{Schwaighofer}, D.~{Raumer}, F.~{Geyer},
  G.~{Carle}, Towards a deeper understanding of tcp bbr congestion control, in:
  2018 IFIP Networking Conference (IFIP Networking) and Workshops, 2018, pp.
  1--9.
\newblock \href {http://dx.doi.org/10.23919/IFIPNetworking.2018.8696830}
  {\path{doi:10.23919/IFIPNetworking.2018.8696830}}.

\bibitem{Jain2018Design}
V.~Jain, V.~Mittal, M.~P. Tahiliani, Design and implementation of tcp bbr in
  ns-3, in: Proceedings of the 10th Workshop on Ns-3, WNS3 '18, ACM, New York,
  NY, USA, 2018, pp. 16--22.
\newblock \href {http://dx.doi.org/10.1145/3199902.3199911}
  {\path{doi:10.1145/3199902.3199911}}.

\bibitem{Ma2017Fairness}
S.~Ma, J.~Jiang, W.~Wang, B.~Li, Fairness of congestion-based congestion
  control: Experimental evaluation and analysis, arXiv preprint
  arXiv:1706.09115.

\bibitem{Wang2019Active}
J.~Wang, Y.~Zheng, Y.~Ni, C.~Xu, F.~Qian, W.~Li, W.~Jiang, Y.~Cheng, Z.~Cheng,
  Y.~Li, X.~Xie, Y.~Sun, Z.~Wang,
  \href{http://doi.acm.org/10.1145/3300061.3300123}{An active-passive
  measurement study of tcp performance over lte on high-speed rails}, in: The
  25th Annual International Conference on Mobile Computing and Networking,
  MobiCom '19, ACM, New York, NY, USA, 2019, pp. 18:1--18:16.
\newblock \href {http://dx.doi.org/10.1145/3300061.3300123}
  {\path{doi:10.1145/3300061.3300123}}.
\newline\urlprefix\url{http://doi.acm.org/10.1145/3300061.3300123}

\bibitem{Ha2008CUBIC}
S.~Ha, I.~Rhee, L.~Xu, Cubic: a new tcp-friendly high-speed tcp variant, ACM
  SIGOPS operating systems review 42~(5) (2008) 64--74.

\bibitem{Misra2000Fluid}
V.~Misra, W.-B. Gong, D.~Towsley, Fluid-based analysis of a network of aqm
  routers supporting tcp flows with an application to red, SIGCOMM Comput.
  Commun. Rev. 30~(4) (2000) 151--160.
\newblock \href {http://dx.doi.org/10.1145/347057.347421}
  {\path{doi:10.1145/347057.347421}}.

\bibitem{Chiu1989Analysis}
D.-M. Chiu, R.~Jain, Analysis of the increase and decrease algorithms for
  congestion avoidance in computer networks, Computer Networks and ISDN Systems
  17~(1) (1989) 1 -- 14.
\newblock \href
  {http://dx.doi.org/https://doi.org/10.1016/0169-7552(89)90019-6}
  {\path{doi:https://doi.org/10.1016/0169-7552(89)90019-6}}.

\bibitem{Kelly1998Rate}
F.~P. Kelly, A.~K. Maulloo, D.~K.~H. Tan, Rate control for communication
  networks: shadow prices, proportional fairness and stability, Journal of the
  Operational Research Society 49~(3) (1998) 237--252.
\newblock \href {http://dx.doi.org/10.1057/palgrave.jors.2600523}
  {\path{doi:10.1057/palgrave.jors.2600523}}.

\bibitem{Low1999Optimization}
S.~H. {Low}, D.~E. {Lapsley}, Optimization flow control. i. basic algorithm and
  convergence, IEEE/ACM Transactions on Networking 7~(6) (1999) 861--874.
\newblock \href {http://dx.doi.org/10.1109/90.811451}
  {\path{doi:10.1109/90.811451}}.

\bibitem{Kelly2003Scalable}
T.~Kelly, Scalable tcp: Improving performance in highspeed wide area networks,
  SIGCOMM Comput. Commun. Rev. 33~(2) (2003) 83--91.
\newblock \href {http://dx.doi.org/10.1145/956981.956989}
  {\path{doi:10.1145/956981.956989}}.

\bibitem{Leith2004H}
D.~Leith, R.~Shorten, H-tcp: Tcp for high-speed and long-distance networks, in:
  Proceedings of PFLDnet, Vol. 2004, 2004.

\bibitem{LisongXu2004Binary}
{Lisong Xu}, K.~{Harfoush}, {Injong Rhee}, Binary increase congestion control
  (bic) for fast long-distance networks, in: IEEE INFOCOM 2004, Vol.~4, 2004,
  pp. 2514--2524 vol.4.
\newblock \href {http://dx.doi.org/10.1109/INFCOM.2004.1354672}
  {\path{doi:10.1109/INFCOM.2004.1354672}}.

\bibitem{Brakmo1995TCP}
L.~S. Brakmo, L.~L. Peterson, Tcp vegas: End to end congestion avoidance on a
  global internet, IEEE Journal on selected Areas in communications 13~(8)
  (1995) 1465--1480.

\bibitem{Wei2006FAST}
D.~X. Wei, C.~Jin, S.~H. Low, S.~Hegde, Fast tcp: motivation, architecture,
  algorithms, performance, IEEE/ACM transactions on Networking 14~(6) (2006)
  1246--1259.

\bibitem{Kuzmanovic2003TCP}
A.~{Kuzmanovic}, E.~W. {Knightly}, Tcp-lp: a distributed algorithm for low
  priority data transfer, in: IEEE INFOCOM 2003. Twenty-second Annual Joint
  Conference of the IEEE Computer and Communications Societies (IEEE Cat.
  No.03CH37428), Vol.~3, 2003, pp. 1691--1701 vol.3.
\newblock \href {http://dx.doi.org/10.1109/INFCOM.2003.1209192}
  {\path{doi:10.1109/INFCOM.2003.1209192}}.

\bibitem{Rossi2010LEDBAT}
D.~{Rossi}, C.~{Testa}, S.~{Valenti}, L.~{Muscariello}, Ledbat: The new
  bittorrent congestion control protocol, in: 2010 Proceedings of 19th
  International Conference on Computer Communications and Networks, 2010, pp.
  1--6.
\newblock \href {http://dx.doi.org/10.1109/ICCCN.2010.5560080}
  {\path{doi:10.1109/ICCCN.2010.5560080}}.

\bibitem{ChengPengFu2003TCP}
{Cheng Peng Fu}, S.~C. {Liew}, Tcp veno: Tcp enhancement for transmission over
  wireless access networks, IEEE Journal on Selected Areas in Communications
  21~(2) (2003) 216--228.
\newblock \href {http://dx.doi.org/10.1109/JSAC.2002.807336}
  {\path{doi:10.1109/JSAC.2002.807336}}.

\bibitem{Liu2008TCP}
S.~Liu, T.~Başar, R.~Srikant, Tcp-illinois: A loss- and delay-based congestion
  control algorithm for high-speed networks, Performance Evaluation 65~(6)
  (2008) 417 -- 440, innovative Performance Evaluation Methodologies and Tools:
  Selected Papers from ValueTools 2006.
\newblock \href {http://dx.doi.org/https://doi.org/10.1016/j.peva.2007.12.007}
  {\path{doi:https://doi.org/10.1016/j.peva.2007.12.007}}.

\bibitem{Tan2006Compound}
K.~{Tan}, J.~{Song}, Q.~{Zhang}, M.~{Sridharan}, A compound tcp approach for
  high-speed and long distance networks, in: Proceedings IEEE INFOCOM 2006.
  25TH IEEE International Conference on Computer Communications, 2006, pp.
  1--12.
\newblock \href {http://dx.doi.org/10.1109/INFOCOM.2006.188}
  {\path{doi:10.1109/INFOCOM.2006.188}}.

\bibitem{David2010Improved}
A.~H. {David}, G.~{Armitage}, Improved coexistence and loss tolerance for delay
  based tcp congestion control, in: IEEE Local Computer Network Conference,
  2010, pp. 24--31.
\newblock \href {http://dx.doi.org/10.1109/LCN.2010.5735714}
  {\path{doi:10.1109/LCN.2010.5735714}}.

\bibitem{Hayes2011Revisiting}
D.~A. Hayes, G.~Armitage, Revisiting tcp congestion control using delay
  gradients, in: J.~Domingo-Pascual, P.~Manzoni, S.~Palazzo, A.~Pont,
  C.~Scoglio (Eds.), NETWORKING 2011, Springer Berlin Heidelberg, Berlin,
  Heidelberg, 2011, pp. 328--341.

\bibitem{Winstein2013Stochastic}
K.~Winstein, A.~Sivaraman, H.~Balakrishnan, Stochastic forecasts achieve high
  throughput and low delay over cellular networks, in: Presented as part of the
  10th USENIX Symposium on Networked Systems Design and Implementation (NSDI
  13), 2013, pp. 459--471.

\bibitem{Dong2015PCC}
M.~Dong, Q.~Li, D.~Zarchy, P.~B. Godfrey, M.~Schapira, Pcc: Re-architecting
  congestion control for consistent high performance, in: 12th USENIX Symposium
  on Networked Systems Design and Implementation (NSDI 15), 2015, pp. 395--408.

\bibitem{Arun2018Copa}
V.~Arun, H.~Balakrishnan, Copa: Practical delay-based congestion control for
  the internet, in: 15th USENIX Symposium on Networked Systems Design and
  Implementation (NSDI 18), 2018, pp. 329--342.

\bibitem{Zaki2015Adaptive}
Y.~Zaki, T.~P\"{o}tsch, J.~Chen, L.~Subramanian, C.~G\"{o}rg, Adaptive
  congestion control for unpredictable cellular networks, SIGCOMM Comput.
  Commun. Rev. 45~(4) (2015) 509--522.
\newblock \href {http://dx.doi.org/10.1145/2829988.2787498}
  {\path{doi:10.1145/2829988.2787498}}.

\bibitem{Park2018ExLL}
S.~Park, J.~Lee, J.~Kim, J.~Lee, S.~Ha, K.~Lee, Exll: An extremely low-latency
  congestion control for mobile cellular networks, in: Proceedings of the 14th
  International Conference on Emerging Networking EXperiments and Technologies,
  CoNEXT '18, ACM, New York, NY, USA, 2018, pp. 307--319.
\newblock \href {http://dx.doi.org/10.1145/3281411.3281430}
  {\path{doi:10.1145/3281411.3281430}}.

\bibitem{Abbasloo2019C2TCP}
S.~{Abbasloo}, Y.~{Xu}, H.~J. {Chao}, C2tcp: A flexible cellular tcp to meet
  stringent delay requirements, IEEE Journal on Selected Areas in
  Communications 37~(4) (2019) 918--932.
\newblock \href {http://dx.doi.org/10.1109/JSAC.2019.2898758}
  {\path{doi:10.1109/JSAC.2019.2898758}}.

\bibitem{Alizadeh2010Data}
M.~Alizadeh, A.~Greenberg, D.~A. Maltz, J.~Padhye, P.~Patel, B.~Prabhakar,
  S.~Sengupta, M.~Sridharan,
  \href{http://dl.acm.org/citation.cfm?id=2043164.1851192}{Data center tcp
  (dctcp)}, SIGCOMM Comput. Commun. Rev. 41~(4) (2010) --.
\newline\urlprefix\url{http://dl.acm.org/citation.cfm?id=2043164.1851192}

\bibitem{Mittal2015TIMELY}
R.~Mittal, V.~T. Lam, N.~Dukkipati, E.~Blem, H.~Wassel, M.~Ghobadi, A.~Vahdat,
  Y.~Wang, D.~Wetherall, D.~Zats, Timely: Rtt-based congestion control for the
  datacenter, SIGCOMM Comput. Commun. Rev. 45~(4) (2015) 537--550.
\newblock \href {http://dx.doi.org/10.1145/2829988.2787510}
  {\path{doi:10.1145/2829988.2787510}}.

\bibitem{Winstein2013TCP}
K.~Winstein, H.~Balakrishnan, Tcp ex machina: Computer-generated congestion
  control, SIGCOMM Comput. Commun. Rev. 43~(4) (2013) 123--134.
\newblock \href {http://dx.doi.org/10.1145/2534169.2486020}
  {\path{doi:10.1145/2534169.2486020}}.

\bibitem{Li2018QTCP}
W.~{Li}, F.~{Zhou}, K.~R. {Chowdhury}, W.~M. {Meleis}, Qtcp: Adaptive
  congestion control with reinforcement learning, IEEE Transactions on Network
  Science and Engineering (2018) 1--1\href
  {http://dx.doi.org/10.1109/TNSE.2018.2835758}
  {\path{doi:10.1109/TNSE.2018.2835758}}.

\bibitem{Xiao2019TCP}
K.~{Xiao}, S.~{Mao}, J.~K. {Tugnait}, Tcp-drinc: Smart congestion control based
  on deep reinforcement learning, IEEE Access 7 (2019) 11892--11904.
\newblock \href {http://dx.doi.org/10.1109/ACCESS.2019.2892046}
  {\path{doi:10.1109/ACCESS.2019.2892046}}.

\bibitem{Carlucci2017Congestion}
G.~Carlucci, L.~De~Cicco, S.~Holmer, S.~Mascolo, G.~Carlucci, L.~De~Cicco,
  S.~Holmer, S.~Mascolo, Congestion control for web real-time communication,
  IEEE/ACM Trans. Netw. 25~(5) (2017) 2629--2642.
\newblock \href {http://dx.doi.org/10.1109/TNET.2017.2703615}
  {\path{doi:10.1109/TNET.2017.2703615}}.

\bibitem{Zhu2018NADA}
X.~Zhu, R.~Pan, M.~Ramalho, S.~de~la Cruz, C.~Ganzhorn, P.~Jones,
  S.~D’Aronco,
  \href{https://tools.ietf.org/html/draft-ietf-rmcat-nada-07}{Nada: A unified
  congestion control scheme for real-time media}, Internet-draft, Internet
  Engineering Task Force, work in Progress (2018).
\newline\urlprefix\url{https://tools.ietf.org/html/draft-ietf-rmcat-nada-07}

\bibitem{Johansson2017Self}
I.~Johansson, Z.~Sarker,
  \href{https://www.rfc-editor.org/rfc/rfc8298.txt}{Self-clocked rate
  adaptation for multimedia}, RFC 8298, RFC Editor (Dec 2017).
\newline\urlprefix\url{https://www.rfc-editor.org/rfc/rfc8298.txt}

\bibitem{Kleinrock1979Power}
L.~Kleinrock, Power and deterministic rules of thumb for probabilistic problems
  in computer communications, in: Proceedings of the International Conference
  on Communications, Vol.~43, 1979, pp. 1--10.

\bibitem{Jaffe1981Flow}
J.~{Jaffe}, Flow control power is nondecentralizable, IEEE Transactions on
  Communications 29~(9) (1981) 1301--1306.
\newblock \href {http://dx.doi.org/10.1109/TCOM.1981.1095152}
  {\path{doi:10.1109/TCOM.1981.1095152}}.

\bibitem{Cardwell2019BBR}
N.~Cardwell, Y.~Cheng, S.~H. Yeganeh, I.~Swett, V.~Jacobson,
  \href{https://datatracker.ietf.org/meeting/104/materials/slides-104-iccrg-an-update-on-bbr-00}{Bbr
  v2 a model-based congestion control}, 2019.
\newline\urlprefix\url{https://datatracker.ietf.org/meeting/104/materials/slides-104-iccrg-an-update-on-bbr-00}

\bibitem{Jain1984quantitative}
R.~K. Jain, D.-M.~W. Chiu, W.~R. Hawe, A quantitative measure of fairness and
  discrimination for resource allocation in shared computer systems, Eastern
  Research Laboratory, Digital Equipment Corporation, Hudson, MA.

\bibitem{Atxutegi2018Use}
E.~{Atxutegi}, F.~{Liberal}, H.~K. {Haile}, K.~{Grinnemo}, A.~{Brunstrom},
  A.~{Arvidsson}, On the use of tcp bbr in cellular networks, IEEE
  Communications Magazine 56~(3) (2018) 172--179.
\newblock \href {http://dx.doi.org/10.1109/MCOM.2018.1700725}
  {\path{doi:10.1109/MCOM.2018.1700725}}.

\end{thebibliography}
\end{document}